\documentclass[acmtog, nonacm]{acmart}
\acmSubmissionID{1151}

\usepackage{booktabs} % For formal tables
\usepackage[table]{xcolor} % For table row colors
\usepackage[ruled]{algorithm2e} % For algorithms
\usepackage{comment}

\usepackage[capitalise]{cleveref}
\usepackage{multirow}
\usepackage{makecell}
\usepackage{pifont}
\usepackage{xspace}

\usepackage{kotex}

\renewcommand{\algorithmcfname}{ALGORITHM}
\SetAlFnt{\small}
\SetAlCapFnt{\small}
\SetAlCapNameFnt{\small}
\SetAlCapHSkip{0pt}

\acmJournal{TOG}
\acmYear{2026}
\acmMonth{1}
\copyrightyear{2026}

\newcommand{\ours}{SceneFrom3D\xspace}

\begin{document}

% Title portion

% \title{\ours{}: 3D Scene Generation via Geometry Prompting}
\title{\ours{}: Geometry-Conditioned Outdoor 3D Scene Generation via View Scheduling with Object-Level Control}

% High-resolution with large content expansion ratios.

% AUTHOR INFORMATION
\author{Geonung Kim}
\affiliation{
  \institution{POSTECH}
  \country{Republic of Korea}
}
\email{k2woong92@postech.ac.kr}

\author{Jeongeun Park}
\affiliation{
  \institution{POSTECH}
  \country{Republic of Korea}
}
\email{koyy001@postech.ac.kr}

\author{Nuri Ryu}
\affiliation{
  \institution{POSTECH}
  \country{Republic of Korea}
}
\email{ryunuri@postech.ac.kr}

\author{Di Liu}
\affiliation{
  \institution{Meta Reality Labs}
  \country{United States of America}
}
\email{lsn33096@gmail.com}

\author{Sunghyun Cho}
\affiliation{
  \institution{POSTECH}
  \country{Republic of Korea}
}
\email{s.cho@postech.ac.kr}

%\author{Gang Zhou}
%\orcid{1234-5678-9012-3456}
%\affiliation{%
%  \institution{College of William and Mary}
%  \streetaddress{104 Jamestown Rd}
%  \city{Williamsburg}
%  \state{VA}
%  \postcode{23185}
%  \country{USA}}
%\email{gang_zhou@wm.edu}
%\author{Valerie B\'eranger}
%\affiliation{%
%  \institution{Inria Paris-Rocquencourt}
%  \city{Rocquencourt}
%  \country{France}
%}
%\email{beranger@inria.fr}
%\author{Aparna Patel}
%\affiliation{%
% \institution{Rajiv Gandhi University}
% \streetaddress{Rono-Hills}
% \city{Doimukh}
% \state{Arunachal Pradesh}
% \country{India}}
%\email{aprna_patel@rguhs.ac.in}
%\author{Huifen Chan}
%\affiliation{%
%  \institution{Tsinghua University}
%  \streetaddress{30 Shuangqing Rd}
%  \city{Haidian Qu}
%  \state{Beijing Shi}
%  \country{China}
%}
%\email{chan0345@tsinghua.edu.cn}
%\author{Ting Yan}
%\affiliation{%
%  \institution{Eaton Innovation Center}
%  \city{Prague}
%  \country{Czech Republic}}
%\email{yanting02@gmail.com}
%\author{Tian He}
%\affiliation{%
%  \institution{University of Virginia}
%  \department{School of Engineering}
%  \city{Charlottesville}
%  \state{VA}
%  \postcode{22903}
%  \country{USA}
%}
%\affiliation{%
%  \institution{University of Minnesota}
%  \country{USA}}
%\email{tinghe@uva.edu}
%\author{Chengdu Huang}
%\author{John A. Stankovic}
%\author{Tarek F. Abdelzaher}
%\affiliation{%
%  \institution{University of Virginia}
%  \department{School of Engineering}
%  \city{Charlottesville}
%  \state{VA}
%  \postcode{22903}
%  \country{USA}
%}
    
%\renewcommand\shortauthors{Zhou, G. et al}

\newcommand{\di}[1]{{\color{magenta} (Di: {#1})}}

%% article.

\begin{abstract}
Geometry-conditioned 3D scene generation enables the creation of 3D environments from user-provided geometry, offering direct control over scene structure and object layout. To generate such 3D scenes, current methods commonly adopt a three-stage design that first defines a view schedule, then synthesizes multi-view observations along the scheduled views, and finally reconstructs a 3D representation from the generated images. However, defining the view schedule becomes a major bottleneck for outdoor scenes, where large, unstructured, and unbounded geometry makes it difficult to obtain views that provide sufficient coverage while supporting stable generation. To address this bottleneck, we present \ours, a framework that automatically schedules views from outdoor input geometries. \ours constructs a directed generation graph whose nodes represent anchor views and whose edges represent interpolation trajectories, defining which views to synthesize, which view pairs to interpolate, and in which order generation should proceed. Beyond automatic view scheduling, \ours further improves controllability through object-level conditioning, assigning each object an identity image for appearance guidance and a geometry-adherence parameter for region-wise control over the input geometry. Experiments demonstrate that \ours achieves state-of-the-art geometry-conditioned outdoor 3D scene generation, producing high-quality scenes with controllable object appearance and geometry adherence. 
Project page: \href{https://kimgeonung.github.io/SceneFrom3D}{\textcolor{purple}{kimgeonung.github.io/SceneFrom3D}}.
\end{abstract}

\begin{teaserfigure}
   \centering
   \includegraphics[width=1.0\textwidth]{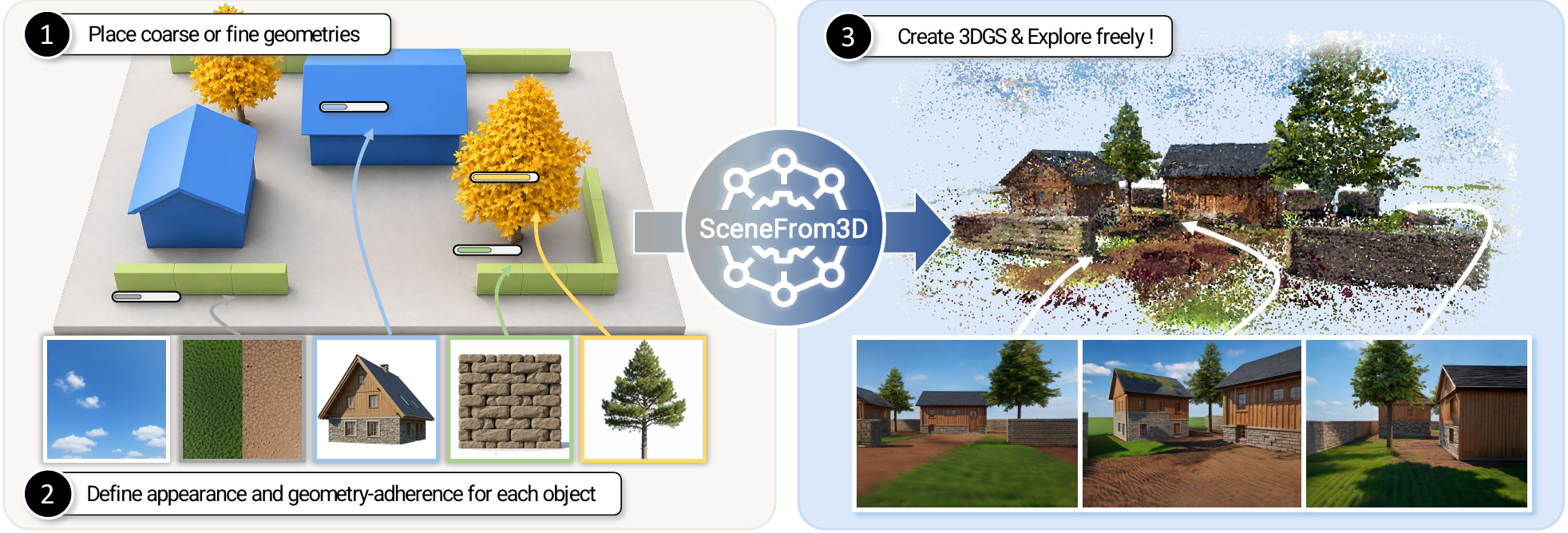}
   \vspace{-4mm}
   \caption[teaser]{
   Overview of \ours. Given coarse or fine object geometries with per-object appearance and geometry-adherence conditions, \ours generates a 3DGS scene that can be rendered and explored from arbitrary viewpoints.
   }
   \label{fig:teaser}
\end{teaserfigure}
\maketitle

\section{Introduction}
3D scene generation aims to synthesize a spatially consistent 3D representation of a scene, typically represented by Neural Radiance Fields (NeRF)~\cite{nerf} or 3D Gaussian Splatting (3DGS)~\cite{3dgs}.
These representations allow rendering from arbitrary viewpoints while maintaining view consistency, enabling users to navigate and explore the generated scene freely.
This capability makes them well-suited for immersive and interactive experiences, leading to broad applications in virtual reality, gaming, simulation, robotics, and autonomous driving.

Recent research in 3D scene generation has explored various input modalities, including text prompts~\cite{text2room}, images~\cite{realmdreamer}, scene graphs~\cite{liu2025controllable}, and geometric conditions~\cite{controlroom3d}.
Among these, geometric conditions provide explicit spatial priors for scene composition in the form of meshes or semantic layouts.
Such conditions significantly reduce the ambiguity of the generated scene, leading to more reliable results while providing users with practical control over the generation process.

Since paired data between geometry and high-quality 3D scenes is scarce and difficult to acquire, recent state-of-the-art methods commonly decompose geometry-conditioned 3D scene generation into a three-stage pipeline~\cite{worldmesh,spatialgen,scenecraft,videofrom3d}. Specifically, given geometric conditions as input, the first stage defines camera viewpoints or trajectories that cover the underlying scene. The second stage employs diffusion models to synthesize geometry-conditioned multi-view observations. The final stage uses the synthesized observations to optimize a 3D representation, such as NeRF or 3DGS. By separating multi-view generation from 3D reconstruction, this pipeline can exploit strong image or video diffusion priors while avoiding the need to train a direct mapping from geometry to complete 3D scenes.

The success of this three-stage pipeline critically depends on the first stage of camera viewpoint definition. The selected viewpoints must provide sufficient coverage of the input geometry so that the generated observations support complete 3D reconstruction. At the same time, they must be suitable for diffusion-based generation, with an appropriate number of views and well-distributed viewing directions. Existing methods obtain such viewpoints under restrictive assumptions. Indoor methods exploit bounded and structured room layouts, where rule-based camera placement can be effective~\cite{worldmesh,spatialgen,scenecraft}. Other methods assume that camera trajectories are provided as input~\cite{videofrom3d}. These assumptions are difficult to satisfy for arbitrary outdoor geometry, which often has unstructured layout, open boundaries, and no canonical camera range.
Consequently, existing pipelines lack a general mechanism for automatically producing view schedules for arbitrary outdoor geometry.
% As a result, automatic view scheduling for geometry-conditioned outdoor 3D scene generation remains largely underexplored.

In this paper, we propose \ours, a geometry-conditioned framework for outdoor 3D scene generation that automatically produces the view schedule required by this three-stage pipeline. Given arbitrary input geometry, \ours constructs a directed generation graph whose nodes represent anchor views and whose edges represent interpolation trajectories between anchor-view pairs. Our view scheduling algorithm first estimates a compact set of anchor views that covers the input geometry, then connects suitable view pairs for interpolation, and finally orients the graph to define generation dependencies. The graph determines which views to synthesize, which view pairs to interpolate, and in which order multi-view synthesis should proceed, enabling scalable generation in large and unstructured outdoor scenes without predefined camera paths.

In addition to view scheduling, \ours also improves the controllability of generated scenes. Previous geometry-conditioned methods mainly use input geometry to control structural or semantic layout, but provide limited control over the appearance of individual objects and how closely each object follows its input geometry. \ours addresses this limitation through object level conditioning. Each object is assigned an identity image to guide its appearance, while the strength of geometric conditioning is adjusted for each region to control geometry adherence. This design enables both flexible control over global scene structure and precise object-level appearance with adherence to the input geometry.
Together, automatic view scheduling and object-level conditioning allow \ours to generate outdoor 3DGS scenes from input geometries with object-level controllability, as shown in \cref{fig:teaser}.
% Together, automatic view scheduling and object-level conditioning enable \ours to generate controllable outdoor 3DGS scenes from input geometries with object-level appearance and adherence control

Our contributions are summarized as follows:
\begin{itemize}
    \item We propose \ours, the first geometry-conditioned framework for outdoor 3D scene generation that does not require explicit camera trajectories as input.
    \item We introduce an automatic view scheduling algorithm that selects anchor views, interpolation trajectories, and generation order from arbitrary outdoor scene geometry.
    \item 
    We introduce an object-level conditioning mechanism that adds control over object specific appearance and region wise adherence to input geometry.
    \item Extensive experiments demonstrate that \ours generates high quality 3D scenes in complex and unstructured outdoor environments.
\end{itemize}
\section{Related Work}
\paragraph{Geometry-guided 3D generation.}
Early geometry-guided 3D generation methods primarily focused on single objects using Score Distillation Sampling (SDS)~\cite{dreamfusion,fantasia3d,mvedit,coin3d,latentnerf} or multi-view image generation under forward-facing view assumptions~\cite{zero123++,phidias, ryu2023pop3d}. Subsequent works extended these approaches to scene-level generation conditioned on untextured meshes~\cite{worldmesh,mvdiffusion} or semantic layouts~\cite{controlroom3d,roomtex,spatialgen,scenecraft,wang2025worldgentexttraversableinteractive}. However, many of these methods target indoor scenes, where bounded and structured layouts make heuristic camera placement effective. A few recent methods handle outdoor scenes~\cite{urbanarchitect,scenecraft,videofrom3d}, but they assume user-specified or externally provided camera trajectories. Designing such camera trajectories is difficult for large and unstructured scenes, since inefficient placement increases generation cost while incomplete coverage can leave important regions unobserved. In contrast, \ours automatically schedules anchor views and interpolation trajectories from the input geometry, enabling scalable geometry-conditioned generation for outdoor scenes.

\begin{figure*}[t]
    \centering
    \includegraphics[width=1.0\textwidth]{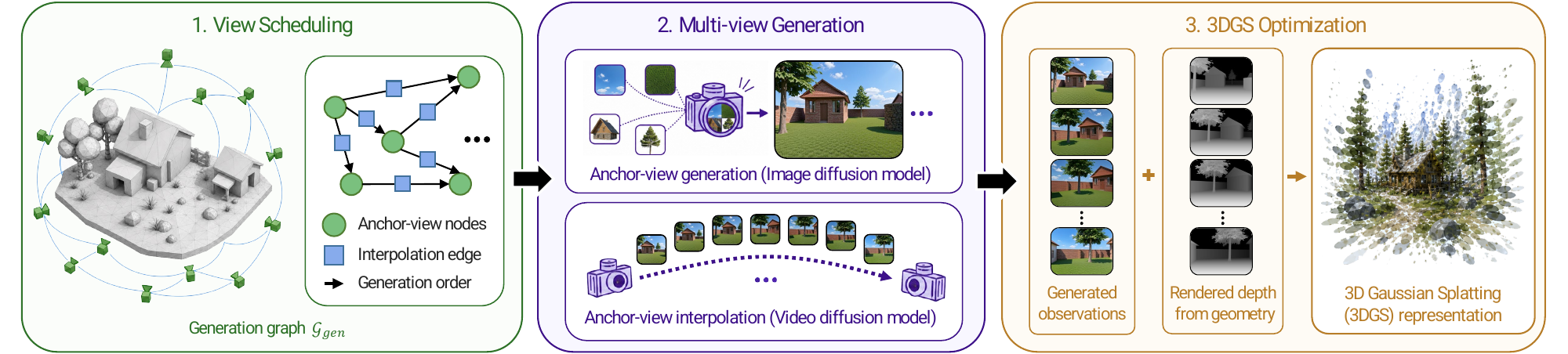}
    \vspace{-5mm}
    \caption{
    % Overview of \ours. 
    % Given object-level geometry prompts with identity images and geometry-adherence parameters, \ours performs automatic view scheduling to construct a directed generation graph. The graph specifies anchor views, interpolation trajectories, and generation order for multi-view generation, and the generated posed observations are optimized into a 3DGS representation.
    Framework overview.
    \ours first performs view scheduling from the input mesh to construct a directed generation graph. It then generates multi-view images by using identity images and geometry-adherence parameters for anchor-view generation and video diffusion for anchor-view interpolation. 
    Finally, the generated posed observations and rendered depth maps from the input mesh are jointly used to optimize a 3DGS representation. 
    }
    \label{fig:framework}
\end{figure*}

\paragraph{View scheduling.}
View scheduling has also been widely studied in robotics and vision, most notably as next-best-view (NBV) planning, which selects the most informative next viewpoint for tasks such as active object recognition~\cite{dickinson1997active,roy2000isolated}, inspection~\cite{tarbox1995planning,trucco2002model}, and 3D reconstruction~\cite{tarbox1995planning,trucco2002model}. Aerial path planning further extends this idea by optimizing camera trajectories and viewpoints for large-scale reconstruction using drones~\cite{continuous,submodular,smith2018aerial,zhou2020offsite,liu2022learning,tang2025aerial}. Although these methods also determine viewpoints and trajectories from geometric information, their objectives and constraints differ from ours. They target reconstruction under physical acquisition constraints, where a camera must follow a single continuous path. In our setting, no physical camera traverses the scene, and generation can branch from any synthesized anchor view. A single global path is therefore inefficient, so \ours formulates view scheduling as graph construction and plans multiple local interpolation trajectories tailored to geometry-conditioned outdoor 3D scene generation.

% \paragraph{View Scheduling}
% View scheduling has also been studied outside generative 3D scene synthesis, most notably through next-best-view (NBV) planning.
% The Next Best View (NBV) problem seeks to select the most informative viewpoint for a given task, such as active object recognition~\cite{dickinson1997active,roy2000isolated}, inspection~\cite{tarbox1995planning,trucco2002model}, and 3D reconstruction~\cite{tarbox1995planning,trucco2002model}, and has been widely studied in robotics and vision. 
% As a specific application of NBV, aerial path planning optimizes camera trajectories and viewpoints for large-scale reconstruction using drones~\cite{continuous,submodular,smith2018aerial,zhou2020offsite,liu2022learning,tang2025aerial,xiong2025aerial}.
% This conceptually aligns with our goal, as it also determines viewpoints and trajectories given proxy geometry, seemingly addressing our problem well.
% However, these methods focus on reconstruction rather than generation, and assume a single continuous trajectory due to the physical constraints of drones, which is highly inefficient in our context.
% To handle this, we formulate view scheduling as a graph problem, enabling effective multi-trajectory planning tailored for outdoor 3D scene generation.

\paragraph{Object-level conditioning.}
Controllable generation has increasingly moved from global conditioning toward spatially localized guidance. In 2D image generation, early text-conditioned diffusion models~\cite{ldm} were extended with structural controls such as depth, edges, and segmentation maps~\cite{controlnet}, as well as region-level conditioning for localized editing and composition~\cite{reco}. A similar trend appears in 3D scene generation, where text-driven methods~\cite{text2room,realmdreamer} have been extended with scene-level structural priors such as layouts, semantic maps, and coarse geometry~\cite{controlroom3d,videofrom3d}. However, existing 3D methods mainly use such conditions to guide the global scene structure, leaving limited control over the appearance of individual objects and the degree to which each object should follow the input geometry. \ours addresses this limitation through object-level conditioning, where each object is associated with an identity image for appearance guidance and a geometry-adherence parameter for region-wise control over the input geometry.

% \paragraph{Fine-grained Controllability}
% \discuss{이 부분을 granularity의 문제로 포장하는 게 좋을 지 모르겠습니다. Granularity로 따지면 sparse<->fine이 생각납니다. 하지만 이 경우 sparse control만 가능하다가 이제 정말 세밀한 geometry control도 가능해지는게 아닌가 하는 생각이 들 것 같아요.}
% \kkw{-grain 라는 표현 자체가 주는 느낌이 문제라고 생각하는것인지?/ 두 번째 문장은 이해 못했음. rough control -> fine control로 가는거 맞음}
% Fine-grained control is essential in visual content generation, as users often need precise manipulation over both structure and appearance to achieve specific outcomes.
% For example, in 2D image generation, conditioning has evolved from global text prompts~\cite{ldm} to structure guidance~\cite{controlnet} and even region-controlled generation~\cite{reco}.
% This trend is also reflected in 3D scene generation, which has progressed from text-driven methods~\cite{text2room,realmdreamer} to approaches incorporating structural priors, such as layouts or coarse geometry~\cite{controlroom3d,videofrom3d}.
% However, existing methods largely lack object-level controllability, limiting their ability to specify detailed appearance and structure of individual elements within a scene.
% To address this limitation, we propose a framework that enables fine-grained, object-level control over both appearance and geometric structure in 3D scene generation.

\section{Methods}
\label{sec:methods}
\ours aims to generate an outdoor 3D scene from object-level geometry prompts. Let $\mathcal{O} \equiv \{1,\dots,N\}$ denote the set of object indices in the scene. The input is a set of attributed object meshes,
\begin{align}
    \mathcal{M} \equiv \{\mathbf{m}_o\}_{o \in \mathcal{O}},
    \qquad
    \mathbf{m}_o \equiv (M_o, I_o, \alpha_o),
\end{align}
where $M_o$ denotes the input geometry of object $o$, $I_o$ denotes an identity image specifying its target appearance, and $\alpha_o \in [0,1]$ denotes a geometry-adherence parameter. Given $\mathcal{M}$, \ours synthesizes a 3DGS representation of the output scene. 

\subsection{Overview}
\label{sec:methods_overview}
\cref{fig:framework} illustrates the overall framework of \ours. First, \ours performs automatic view scheduling over the input geometry. The scheduler constructs a directed generation graph, where nodes represent anchor views and directed edges represent interpolation trajectories with generation dependencies. This graph determines the anchor views, interpolation pairs, and generation order.

Second, \ours generates multi-view observations according to the generation graph. We follow the generation strategy of VideoFrom3D~\cite{videofrom3d}, which combines image-based anchor-view synthesis with video-based view interpolation for geometry-conditioned scene generation. Unlike VideoFrom3D, which relies on given camera trajectories, \ours uses the generation graph to organize the same generation process. During anchor-view synthesis, rendered geometry, object identity images, and region-wise conditioning strengths guide appearance synthesis and adherence to the input geometry.

Finally, \ours optimizes a 3DGS representation from the generated multi-view observations and mesh-rendered depth maps. The generated images provide posed appearance supervision, while the depth maps encourage alignment with the input geometry. We describe the three stages in detail in the following subsections.

\begin{figure*}[t]
    \centering
    \includegraphics[width=\textwidth]{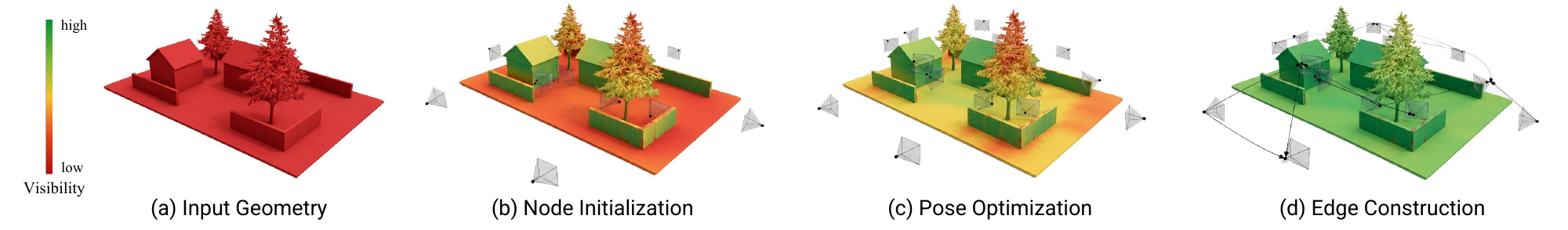}
    \vspace{-5mm}
    \caption{
    Visualization of the view scheduling process.
    (a) The input geometry is color-coded by surface-point visibility.
    (b) Visibility-guided node initialization selects a compact set of anchor views that broadly cover the input geometry.
    (c) Pose optimization refines the anchor-view poses to improve both surface coverage and generation suitability.
    (d) Edge construction links suitable anchor-view pairs with interpolation trajectories.
    }
    \label{fig:view_schedule}
\end{figure*}

\subsection{View scheduling}
\label{sec:methods_schedule}
The view-scheduling stage constructs a directed generation graph
\begin{align}
    \mathcal{G}_{\mathrm{gen}}
    \equiv
    (\mathcal{V}, \mathcal{E}_{\rightarrow}),
\end{align}
where $\mathcal{V} \equiv \{v_i\}_{i=1}^{N_{\mathrm{v}}}$ is the set of anchor-view nodes, $v_i$ denotes the $i$-th anchor view, $N_{\mathrm{v}}$ is the number of anchor views, and $\mathcal{E}_{\rightarrow}$ is the set of directed interpolation edges without self-loops. Each edge $(v_i,v_k) \in \mathcal{E}_{\rightarrow}$ represents an interpolation trajectory from $v_i$ to $v_k$ and specifies that $v_i$ is generated before $v_k$. View scheduling proceeds through node construction, edge construction, and direction construction, which we describe in order below.

\subsubsection{Node construction}
% The objective of node construction is to find a compact anchor-view set $\mathcal{V}$ that broadly covers the input geometry with a minimal number of cameras. 
The objective of node construction is to find a compact anchor-view set $\mathcal{V}$ that broadly covers the input geometry with minimal cameras. 
This compactness is important because intermediate regions are later densely completed through video interpolation, so excessive anchor views provide limited benefit while increasing computation.
In addition, as discussed in VideoFrom3D~\cite{videofrom3d}, placing anchor views too closely may worsen multi-view inconsistency, since stochastic diffusion sampling can produce conflicting details in shared observed regions.

% To construct such a set, we define a visibility measure over sampled surface points of the input meshes, which estimates how well candidate cameras observe the input geometry. We use this measure in two steps. We first initialize anchor views through visibility-guided densification, which places cameras to improve coverage of the input geometry. We then optimize the initial camera poses to further improve coverage while preserving generation-friendly viewpoints. We next describe the visibility measure, followed by node initialization and pose optimization.

% To construct such a set, we define a visibility measure over sampled surface points of the input meshes, which estimates how well candidate cameras observe the input geometry. Based on this measure, we initialize anchor views through visibility-guided densification to improve coverage, and then optimize the initial camera poses to further improve coverage while preserving generation-friendly viewpoints. We next describe the visibility measure, node initialization, and pose optimization.

To construct such a set, we define a visibility measure over sampled surface points of the input meshes to estimate how well candidate cameras observe the input geometry.
Using this measure, we initialize anchor views through visibility-guided densification and then optimize their poses to improve coverage while preserving generation-friendly viewpoints.
We next describe the visibility measure, node initialization, and pose optimization.

\paragraph{Visibility measure.}
To evaluate how well the anchor views cover the input geometry, we uniformly sample surface points from the input meshes as $\mathcal{P} \equiv \{(\mathbf{p}_n,\mathbf{n}_n)\}_{n=1}^{N_{\mathrm{p}}}$, where $\mathbf{p}_n$ and $\mathbf{n}_n$ denote the 3D position and outward normal of the $n$-th sample, respectively, and $N_{\mathrm{p}}$ denotes the number of surface samples.
We evaluate coverage by relating these samples to each anchor-view camera.
Each anchor view $v_i \in \mathcal{V}$ is associated with a camera center $\mathbf{q}_i \in \mathbb{R}^{3}$ and a world-to-camera rotation $\mathbf{R}_i \in \mathrm{SO}(3)$, which together define the world-to-camera transform for view $v_i$.

For an anchor view $v_i$ and a surface sample $\mathbf{p}_n$, we express the sample in the coordinate system of camera $v_i$ as
\begin{align}
    (x^c_{i,n}, y^c_{i,n}, z^c_{i,n})^\top
    \equiv
    \mathbf{R}_i(\mathbf{p}_n-\mathbf{q}_i),
\end{align}
where the superscript $c$ denotes camera coordinates and the positive $z^c$ axis points forward.
We also define the camera-to-sample distance and ray direction as
\begin{align}
    r_{i,n} \equiv \|\mathbf{p}_n-\mathbf{q}_i\|_2,
    \qquad
    \boldsymbol{\omega}_{i,n} \equiv \frac{\mathbf{p}_n-\mathbf{q}_i}{r_{i,n}}.
\end{align}

% For an anchor view $v_i$ and a surface sample $\mathbf{p}_n$, we express the sample in the coordinate system of camera $v_i$ as
% \begin{align}
%     (x^c_{i,n}, y^c_{i,n}, z^c_{i,n})^\top
%     \equiv
%     \mathbf{R}_i(\mathbf{p}_n-\mathbf{q}_i),
% \end{align}
% where the superscript $c$ denotes camera coordinates and the positive $z^c$ axis points forward. We also define
% \begin{align}
%     r_{i,n} \equiv \|\mathbf{p}_n-\mathbf{q}_i\|_2,
%     \qquad
%     \boldsymbol{\omega}_{i,n} \equiv \frac{\mathbf{p}_n-\mathbf{q}_i}{r_{i,n}}.
% \end{align}
% where $r_{i,n}$ is the distance from the camera center to the surface sample and $\boldsymbol{\omega}_{i,n}$ is the ray direction from the camera center to the surface sample.

% For anchor view $v_i$ and surface sample $\mathbf{p}_n$, we define a soft visibility score as
Using these geometric quantities, we define the soft visibility score between anchor view $v_i$ and surface sample $\mathbf{p}_n$ as
\begin{align}
    V_{i,n}
    =
    V^{\mathrm{fov}}_{i,n}
    \cdot
    V^{\mathrm{dist}}_{i,n}
    \cdot
    V^{\mathrm{front}}_{i,n}
    \cdot
    M^{\mathrm{occ}}_{i,n}.
\end{align}
This score measures whether a surface sample is useful for view scheduling by combining four criteria. 
The sample should lie inside the camera field of view, be observed from an appropriate distance, face the camera, and remain unoccluded by the input geometry.

The field-of-view term encourages samples to lie inside the camera frustum.
Since all anchor views share fixed horizontal and vertical field-of-view angles, denoted by $\mathrm{FOV}_x$ and $\mathrm{FOV}_y$, we first compute the angular offsets of each sample in camera coordinates as $\gamma^{x}_{i,n} \equiv \operatorname{atan2}(x^c_{i,n}, z^c_{i,n})$ and $\gamma^{y}_{i,n} \equiv \operatorname{atan2}(y^c_{i,n}, z^c_{i,n})$.
We then define the field-of-view term as
\begin{align}
    V^{\mathrm{fov}}_{i,n}
    =
    \operatorname{sigmoid}
    \left(
        \beta
        \left(
            1-\frac{2|\gamma^{x}_{i,n}|}{\mathrm{FOV}_x}
        \right)
    \right)
    \cdot
    \operatorname{sigmoid}
    \left(
        \beta
        \left(
            1-\frac{2|\gamma^{y}_{i,n}|}{\mathrm{FOV}_y}
        \right)
    \right),
\end{align}
where $\beta$ controls the sharpness of the frustum boundary.

The distance term favors a preferred viewing distance $d_0$,
\begin{align}
    V^{\mathrm{dist}}_{i,n}
    =
    \exp\left(
        -\frac{(r_{i,n}-d_0)^2}{\sigma_{\mathrm{dist}}^2}
    \right),
\end{align}
where $\sigma_{\mathrm{dist}}$ controls the tolerance around $d_0$.

The front-facing term favors surface samples whose normals face the camera,
\begin{align}
    V^{\mathrm{front}}_{i,n}
    =
    (1-\lambda_{\mathrm{front}})
    +
    \lambda_{\mathrm{front}}
    \max\bigl(0,-\langle \boldsymbol{\omega}_{i,n},\mathbf{n}_n\rangle\bigr),
\end{align}
where $\lambda_{\mathrm{front}}$ controls the strength of the front-facing preference, and $\langle \cdot,\cdot \rangle$ denotes the inner product.

Finally, $M^{\mathrm{occ}}_{i,n} \in \{0,1\}$ is a binary visibility mask that equals $1$ when $z^c_{i,n}>0$ and the segment from $\mathbf{q}_i$ to $\mathbf{p}_n$ is not occluded by the input geometry, and equals $0$ otherwise.

Given the anchor-view set $\mathcal{V}$, we aggregate visibility for each surface sample by a soft union,
\begin{align}
    \bar{V}_n
    =
    1
    -
    \prod_{i=1}^{N_{\mathrm{v}}}
    (1-V_{i,n}).
\end{align}
The aggregated score $\bar{V}_n$ becomes high when at least one anchor view observes the surface sample with high visibility. 
% \di{define ${N_{\mathrm{v}}}$ here}

% \paragraph{Node initialization.}
% This stage determines the number of anchor views and their initial camera poses, as shown in \cref{fig:view_schedule}(b). Starting from an empty anchor-view set, we repeatedly identify surface samples whose aggregated visibility is below a threshold $\delta_{\mathrm{vis}}$,
% \begin{align}
%     \mathcal{U}
%     \equiv
%     \{n \in \{1,\dots,N_{\mathrm{p}}\} \mid \bar{V}_{n} < \delta_{\mathrm{vis}}\}.
% \end{align}
% If $\mathcal{U}$ is nonempty, we sample an index $n^{\star}$ from $\mathcal{U}$ and add a new anchor view for the corresponding surface sample. The new camera center is placed along the normal of the selected sample at the preferred distance $d_0$, i.e., $\mathbf{q}_{\mathrm{new}}=\mathbf{p}_{n^{\star}}+d_0\mathbf{n}_{n^{\star}}$. Its yaw and pitch are set so that the camera looks back toward $\mathbf{p}_{n^{\star}}$. We repeat this process until no surface sample remains below the visibility threshold.

\paragraph{Node initialization.}
This stage determines the number of anchor views and their initial camera poses, as shown in \cref{fig:view_schedule}(b).
Starting from an empty anchor-view set, we repeatedly sample a surface point $n^{\star}$ whose aggregated visibility is below the threshold, $\bar{V}_{n^{\star}} < \delta_{\mathrm{vis}}$, and add a new anchor view for it.
The camera center is placed along its normal at the preferred distance $d_0$, i.e., $\mathbf{q}_{\mathrm{new}}=\mathbf{p}_{n^{\star}}+d_0\mathbf{n}_{n^{\star}}$, with yaw and pitch set to look back toward $\mathbf{p}_{n^{\star}}$.
We repeat this process until all surface samples satisfy the visibility threshold.

Since this process targets under-covered surface regions, it can add cameras with negligible visibility contribution or strong overlap with existing views. To obtain a compact initialization, we apply a refinement loop that removes low-contribution cameras, merges camera pairs with high shared visibility, and re-densifies remaining under-covered regions. We repeat this loop until the camera set sufficiently covers the input geometry without redundant views. Additional details are provided in the supplementary material.

% \rnr{이 부분을 좀 줄여서 간단히 언급하기?}
% This densification procedure can still produce redundant cameras because visibility is evaluated at the level of surface samples. For example, a slightly occluded surface sample may trigger an additional camera even when a nearby camera already observes almost the same region. To obtain a compact initialization, we apply a refinement loop after densification. The loop removes cameras with negligible visibility contribution, merges camera pairs with high shared visibility, and adds cameras again for any remaining under-covered regions. We repeat the loop until the camera set converges to a compact initialization that sufficiently covers the input geometry. Additional details are provided in the supplementary material.

\paragraph{Camera pose optimization.}
After initialization, we refine the anchor-view poses while fixing the view count. Specifically, we optimize the camera centers $\{\mathbf{q}_i\}_{i=1}^{N_{\mathrm{v}}}$ and rotations $\{\mathbf{R}_i\}_{i=1}^{N_{\mathrm{v}}}$ to improve coverage while maintaining generation-friendly viewpoints. We minimize
\begin{align}
    \mathcal{L}_{\mathrm{node}}
    =
    \mathcal{L}_{\mathrm{cov}}
    +
    \lambda_{\mathrm{rep}} \mathcal{L}_{\mathrm{rep}}
    +
    \lambda_{\mathrm{tilt}} \mathcal{L}_{\mathrm{tilt}},
\end{align}
where $\lambda_{\mathrm{rep}}$ and $\lambda_{\mathrm{tilt}}$ are weighting coefficients. 

The coverage loss encourages the anchor views to cover all surface samples by maximizing aggregated visibility,
\begin{align}
    \mathcal{L}_{\mathrm{cov}}
    =
    -\frac{1}{N_{\mathrm{p}}}
    \sum_{n=1}^{N_{\mathrm{p}}}
    \bar{V}_{n}.
\end{align}

To prevent cameras from moving into the input geometry, we use a repulsion loss,
\begin{align}
    \mathcal{L}_{\mathrm{rep}}
    =
    \sum_{i=1}^{N_{\mathrm{v}}}
    \left[
        \max
        \left(
            0,
            d_{\mathrm{safe}}
            -
            d(\mathbf{q}_i,\{M_o\}_{o \in \mathcal{O}})
        \right)
    \right]^2,
\end{align}
where $d(\mathbf{q}_i,\{M_o\}_{o \in \mathcal{O}})$ denotes the shortest distance from the camera center $\mathbf{q}_i$ to the input geometry, and $d_{\mathrm{safe}}$ is a safety margin. 

We also regularize camera tilt to prevent cameras from being biased toward densely sampled ground regions.
Let $\mathbf{f}_i$ denote the forward direction of camera $v_i$ in world coordinates. Given the global up direction $\mathbf{g}$, we define a tilt loss,
\begin{align} \mathcal{L}_{\mathrm{tilt}} = \frac{1}{N_{\mathrm{v}}} \sum_{i=1}^{N_{\mathrm{v}}} \langle \mathbf{f}_i,\mathbf{g}\rangle^2. 
\end{align} 

Together, the coverage, repulsion, and tilt terms guide the optimization toward anchor-view poses that comprehensively cover the input geometry, avoid entering the input geometry, and maintain viewing directions suitable for generation, as shown in \cref{fig:view_schedule}(c).

\subsubsection{Edge construction}

After constructing the anchor-view nodes, we connect suitable anchor-view pairs so that video interpolation can provide denser multi-view supervision for 3DGS optimization. 
Since interpolation becomes unreliable when two anchor views are too far apart or observe largely different regions, we connect only pairs with non-negligible shared visibility and collision-free motion.
Specifically, for each pair of distinct anchor views $v_i$ and $v_k$, we compute the shared visibility score $S_{i,k}=\sum_{n=1}^{N_{\mathrm{p}}}\min(V_{i,n},V_{k,n})$.
We connect them if $S_{i,k} > \delta_{\mathrm{shared}}$ and the line segment between their camera centers is collision-free.
For each connected pair of anchor views $v_i$ and $v_k$, we store an interpolation trajectory $\Gamma_{i,k}$ that defines the camera path used to generate intermediate views.
Details of the trajectory construction are provided in the supplementary material.
% Each connection stores an interpolation trajectory $\Gamma_{i,k}$, whose construction is detailed in the supplementary material.

% Although the shared-visibility criterion is effective, it can produce disjoint subgraphs or leaf nodes, which could potentially lead to missing out useful interpolation trajectories between neighboring anchor views. To address this issue, we add a small number of additional interpolation edges by connecting disjoint subgraphs through shortest paths and linking leaf nodes whose camera distance is below a threshold $\delta_{\mathrm{leaf}}$. This step improves graph connectivity while keeping the number of additional edges limited. 

While the visibility-based criterion provides reliable interpolation edges, it can produce disjoint subgraphs or leaf nodes, potentially missing useful trajectories between nearby anchor views. To complement this, we add a small number of distance-based edges by connecting disjoint subgraphs through shortest paths and linking leaf nodes whose camera distance is below $\delta_{\mathrm{leaf}}$. This effectively recovers nearby connections that are beneficial but not captured by the visibility-based criterion.

\subsubsection{Direction construction}
% The final step orients the connections. We compute a canonical ordering of the anchor views and orient each connection from the earlier view to the later view, yielding the directed edge set $\mathcal{E}_{\rightarrow}$ and the final generation graph $\mathcal{G}_{\mathrm{gen}}$. Because every edge follows the same ordering, the graph is acyclic, and topological sorting provides a valid anchor-view generation order. For each directed edge $(v_i,v_k) \in \mathcal{E}_{\rightarrow}$, view $v_i$ is generated before $v_k$ and can serve as a parent conditioning view to improve multi-view consistency during anchor-view synthesis. Additional details are provided in the supplementary material.
The final step orients the connections. Specifically, we use the node indices as the generation order and orient each connection from the lower-index view to the higher-index view, yielding the directed edge set $\mathcal{E}_{\rightarrow}$ and the final generation graph $\mathcal{G}_{\mathrm{gen}}$. Since all edges follow this order, the graph is acyclic, and topological sorting gives a valid anchor-view generation order.
For each directed edge $(v_i,v_k) \in \mathcal{E}_{\rightarrow}$, view $v_i$ is generated before $v_k$ and can serve as a parent conditioning view to improve multi-view consistency during anchor-view synthesis.
%Additional details are provided in the supplementary material.

% \subsubsection{Direction construction}

% The final step assigns generation dependencies to the interpolation connections. A valid generation order is required because an anchor view can use already generated neighboring views as conditioning inputs during multi-view synthesis. We therefore orient the interpolation connections to obtain the directed edge set $\mathcal{E}_{\rightarrow}$ and the final generation graph $\mathcal{G}_{\mathrm{gen}}$.

% Specifically, we compute a canonical ordering of the anchor views and orient each interpolation connection from the earlier view to the later view in this ordering. Since every edge follows the same ordering, the resulting graph is acyclic.
% We then compute a generation order by topological sorting. For any directed edge $(v_i,v_k) \in \mathcal{E}_{\rightarrow}$, view $v_i$ is generated before view $v_k$ and can be used as a parent conditioning view for $v_k$. The directed graph therefore specifies the anchor views, interpolation trajectories, and generation dependencies used in the subsequent multi-view generation stage. Additional details are provided in the supplementary material.

\begin{figure}[t]
    \centering
    \includegraphics[width=\linewidth]{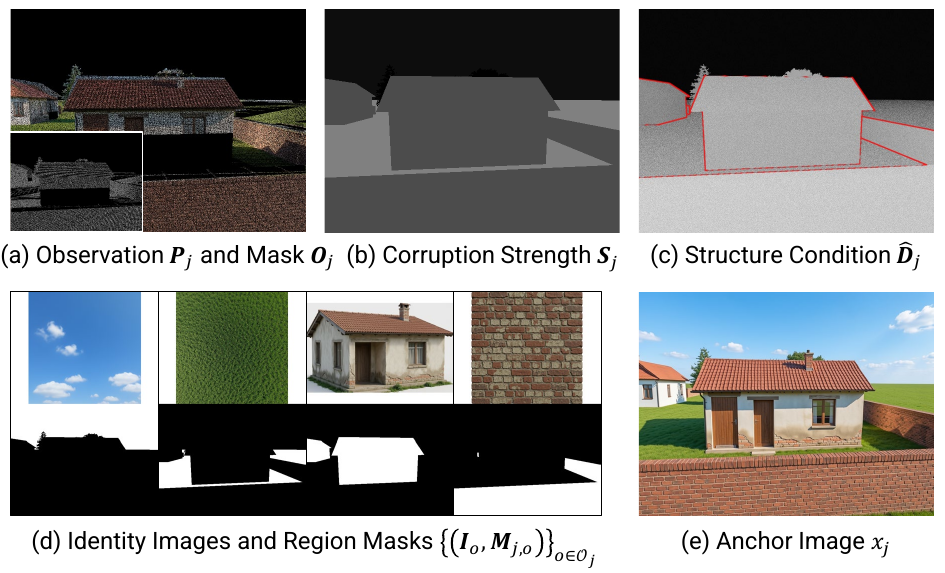}
    \vspace{-5mm}
    \caption{
    % An example of inputs and output for anchor view generation. 
    Example inputs and output for anchor-view generation.
    %The figure shows the partial observation and mask, structure condition, corruption strength map, identity-region conditions, and the generated anchor image for a target view $v_j$. 
    %\di{not sure if this figure provides much info/insights. If we have more space, a cherry-pick figure on  multiview consitency or object w/ w/o hallucination might be good to show}
    }
    \label{fig:mv_gen}
\end{figure}

\subsection{Multi-view generation}
\label{sec:methods_generation}

Given the generation graph $\mathcal{G}_{\mathrm{gen}}$, the multi-view generation stage synthesizes posed observations for 3DGS training. The stage consists of anchor-view generation on graph nodes and view interpolation along graph edges. We describe these two steps in order.

\begin{figure*}[!t]
    \centering
    \includegraphics[width=0.94\textwidth]{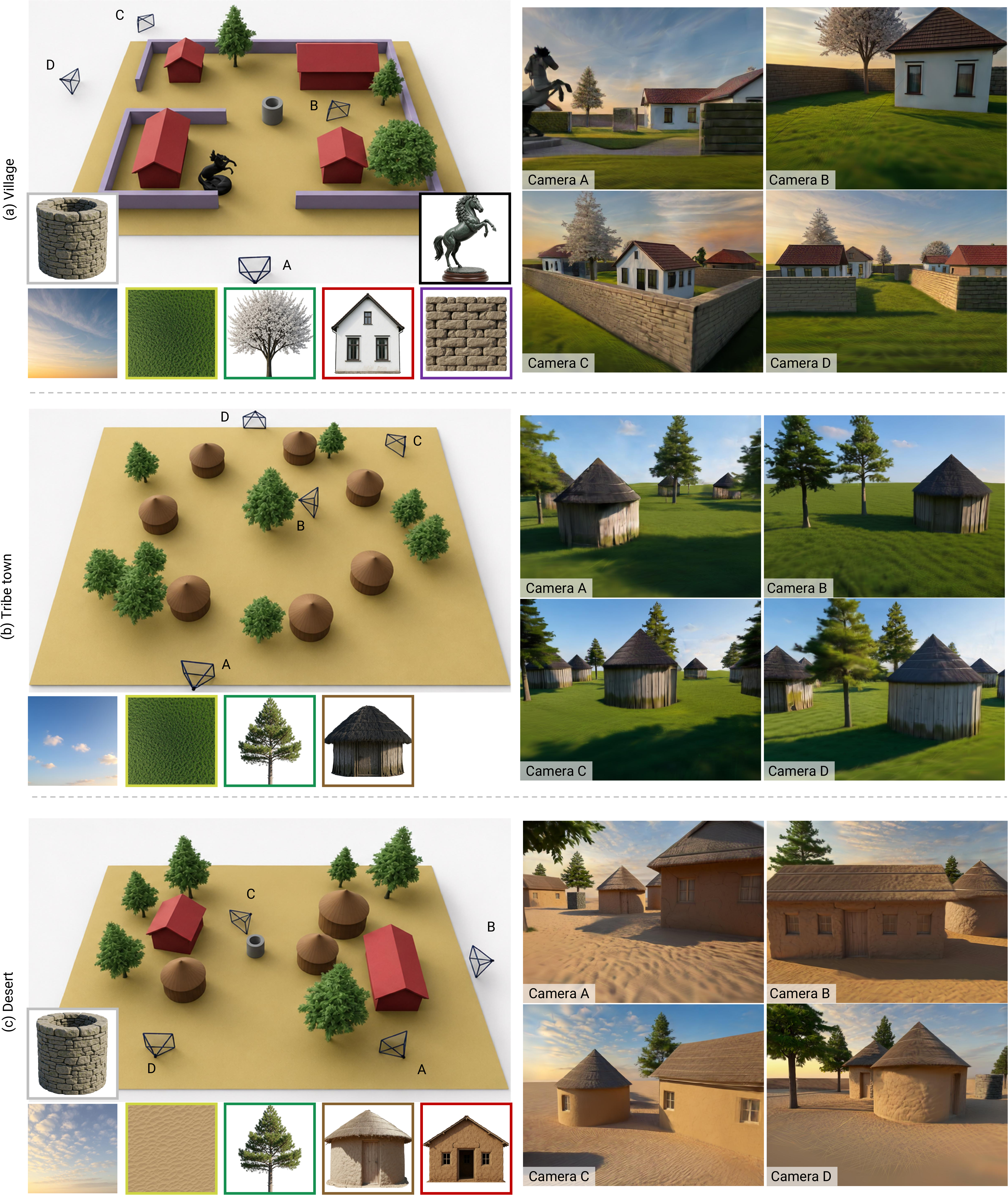}
    \caption{
    Qualitative results across various scenarios.
For each scenario, the figure shows multiple identity images, a 3D model, and rendered images from the camera viewpoints marked in the 3D model.
Colored outlines indicate corresponding elements between the identity images and the 3D model.
The labels A--D denote the camera viewpoints used to generate the rendered images.
    }
    \label{fig:qual}
\end{figure*}

\subsubsection{Anchor-view generation.}
Following the topological order of $\mathcal{G}_{\mathrm{gen}}$, we generate an anchor image for each node $v_j \in \mathcal{V}$.
Let $\mathcal{O}_j \subseteq \mathcal{O}$ denote the set of objects visible from $v_j$.
For each anchor view, we construct a view-specific conditioning input
\begin{align}
    \mathcal{C}_j
    \equiv
    \Bigl(
        \mathbf{P}_j,\,
        \mathbf{O}_j,\,
        \hat{\mathbf{D}}_j,\,
        \mathbf{S}_j,\,
        \{(I_o,\mathbf{M}_{j,o})\}_{o \in \mathcal{O}_j},\,
        \tau_j
    \Bigr),
\end{align}
where $\mathbf{P}_j$ is the partial observation, $\mathbf{O}_j$ is the observation mask, $\hat{\mathbf{D}}_j$ is the structure condition, $\mathbf{S}_j$ is the corruption strength map, $(I_o,\mathbf{M}_{j,o})$ pairs the identity image of object $o$ with its visible-region mask, and $\tau_j$ is the text conditioning prompt.
We feed $\mathcal{C}_j$ to the anchor-view generator to synthesize the anchor image $\mathbf{x}_j$, as visualized in \cref{fig:mv_gen}.

We first construct the partial observation $\mathbf{P}_j$ and observation mask $\mathbf{O}_j$ from the generated parent views of $v_j$ (\cref{fig:mv_gen}(a)).
For each parent view connected to $v_j$ by an incoming edge, we warp its generated image into the target view using correspondences derived from input geometry and camera poses, then merge valid warped pixels to obtain $\mathbf{P}_j$.
The mask $\mathbf{O}_j$ is defined by $\mathbf{O}_j(u)=1$ if pixel $u$ receives a valid warp from any parent view, and $\mathbf{O}_j(u)=0$ otherwise.

We then construct the geometry-based conditions. Specifically, we render the input meshes at the target camera pose to obtain a clean depth map $\mathbf{D}_j$ and object masks $\{\mathbf{M}_{j,o}\}_{o \in \mathcal{O}_j}$. From these masks and the geometry-adherence parameters $\{\alpha_o\}_{o \in \mathcal{O}_j}$, we construct the corruption strength map $\mathbf{S}_j$ (\cref{fig:mv_gen}(b)), setting the strength for object $o$ according to $1-\alpha_o$ so that larger $\alpha_o$ enforces stronger geometric adherence. We sample a Gaussian noise map $\boldsymbol{\epsilon}_j$ and form the corrupted depth map as $\tilde{\mathbf{D}}_j=(1-\mathbf{S}_j)\odot\mathbf{D}_j+\mathbf{S}_j\odot\boldsymbol{\epsilon}_j$, where higher corruption weakens exact geometric guidance. Finally, we overlay boundary cues extracted from the object masks onto $\tilde{\mathbf{D}}_j$ to obtain the structure condition $\hat{\mathbf{D}}_j$ (\cref{fig:mv_gen}(c)), helping align generated object boundaries with the input geometry and reduce warping errors in later partial-observation construction.

% We then construct the geometry-based conditions. Specifically, we render the input meshes at the target camera pose to obtain a clean depth map $\mathbf{D}_j$ and object masks $\{\mathbf{M}_{j,o}\}_{o \in \mathcal{O}_j}$. From these masks and the geometry-adherence parameters $\{\alpha_o\}_{o \in \mathcal{O}_j}$, we construct the corruption strength map $\mathbf{S}_j$, shown in \cref{fig:mv_gen}(b). For pixels belonging to object $o$, the corruption strength is set according to $1-\alpha_o$, so larger $\alpha_o$ preserves the rendered geometry more strongly and smaller $\alpha_o$ relaxes the geometry condition. We sample a Gaussian noise map $\boldsymbol{\epsilon}_j$ and linearly interpolate between the clean depth map and the noise map in a region-wise manner as $\tilde{\mathbf{D}}_j=(1-\mathbf{S}_j)\odot\mathbf{D}_j+\mathbf{S}_j\odot\boldsymbol{\epsilon}_j$. As the corruption strength increases, the structured depth signal is progressively replaced by unstructured noise, which weakens high-frequency geometric cues and reduces the generator's reliance on the exact input geometry. We then overlay boundary cues extracted from the object masks onto $\tilde{\mathbf{D}}_j$, producing the final structure condition $\hat{\mathbf{D}}_j$ shown in \cref{fig:mv_gen}(c). The added boundary cues help align generated object boundaries with the input geometry, reducing warping errors in subsequent partial-observation construction.

The identity-region conditions provide object-specific appearance guidance (\cref{fig:mv_gen}(d)). For each visible object $o \in \mathcal{O}_j$, the pair $(I_o,\mathbf{M}_{j,o})$ specifies that the identity image $I_o$ should guide the appearance of the image region indicated by $\mathbf{M}_{j,o}$. The text prompt $\tau_j$ summarizes the visible object semantics and global scene description based on a predefined rule. Given all conditioning inputs, the anchor-view generator produces the anchor image $\mathbf{x}_j$ (\cref{fig:mv_gen}(e)).

% In implementation, we convert the paired conditioning inputs $(\mathbf{P}_j,\mathbf{O}_j)$, $(\hat{\mathbf{D}}_j,\mathbf{S}_j)$, and $\{(I_o,\mathbf{M}_{j,o})\}_{o \in \mathcal{O}_j}$ into conditioning images through height-wise concatenation. The resulting conditioning images are then combined through image-token concatenation and fed to the generator. We fine-tune the pretrained FLUX.2-klein-9B\footnote{\url{https://huggingface.co/black-forest-labs/FLUX.2-klein-9B}} diffusion model to support this conditioning format. Additional training and inference details are provided in the supplementary material.

In implementation, we convert the paired conditioning inputs $(\mathbf{P}_j,\mathbf{O}_j)$, $(\hat{\mathbf{D}}_j,\mathbf{S}_j)$, and $\{(I_o,\mathbf{M}_{j,o})\}_{o \in \mathcal{O}_j}$ into conditioning images through height-wise concatenation.
The resulting conditioning images are then combined through image-token concatenation and fed to the generator.
We construct a synthetic training dataset and use it to fine-tune the pretrained FLUX.2-klein-9B\footnote{\url{https://huggingface.co/black-forest-labs/FLUX.2-klein-9B}} diffusion model for this conditioning format.
Additional details on dataset construction, training, and inference are provided in the supplementary material.

\subsubsection{Anchor-view interpolation.}
After generating the anchor images, we densify the observation set along the directed edges of $\mathcal{G}_{\mathrm{gen}}$.
For each edge $(v_i,v_j) \in \mathcal{E}_{\rightarrow}$ with its associated trajectory $\Gamma_{i,j}$, we provide the endpoint anchor images $\mathbf{x}_i$ and $\mathbf{x}_j$, along with depth conditions rendered along $\Gamma_{i,j}$, to an off-the-shelf video diffusion model, VACE\footnote{\url{https://huggingface.co/Wan-AI/Wan2.1-VACE-14B}}.
The model synthesizes an interpolation sequence between the two anchors, and the resulting frames provide dense posed observations for the subsequent 3DGS optimization.

\subsection{3DGS optimization}
\label{sec:methods_3dgs}
% Finally, we optimize a 3DGS scene from the generated posed observations. Let $\mathcal{J}$ denote the generated views. For each view $m \in \mathcal{J}$, let $\mathbf{I}_m$ be the generated RGB image and $\mathbf{D}^{\mathrm{mesh}}_m$ be the metric depth rendered from the input meshes at the same pose. We optimize the 3DGS by minimizing
Finally, we optimize a 3DGS scene from the generated posed observations. For each generated view $m \in \mathcal{J}$, let $\mathbf{I}_m$ be its RGB image and $\mathbf{D}^{\mathrm{mesh}}_m$ be the metric depth rendered from the input meshes at the same pose. We optimize the 3DGS by minimizing
% Finally, we optimize a 3DGS scene from the generated posed observations. Let $\mathcal{J}$ denote the index set of generated views, including both anchor views and interpolated views. For each view $m \in \mathcal{J}$, let $\mathbf{I}_m$ denote the generated RGB image and $\mathbf{D}^{\mathrm{mesh}}_m$ denote the metric depth map rendered from the input meshes at the same camera pose. We optimize the 3DGS by minimizing
\begin{align}
    \mathcal{L}_{\mathrm{3DGS}}
    &=
    \frac{1}{|\mathcal{J}|}
    \sum_{m \in \mathcal{J}}
    \Bigl[
    \lambda_{\mathrm{r}}
    \mathcal{L}_{1}
    (\tilde{\mathbf{I}}_m,\mathbf{I}_m)
    +
    \lambda_{\mathrm{s}}
    \mathcal{L}_{\mathrm{DSSIM}}
    (\tilde{\mathbf{I}}_m,\mathbf{I}_m)
    \nonumber\\
    &+
    \lambda_{\mathrm{p}}
    \mathcal{L}_{\mathrm{LPIPS}}
    (\tilde{\mathbf{I}}_m,\mathbf{I}_m)
    +
    \lambda_{\mathrm{d}}
    \|
    \tilde{\mathbf{D}}_m
    -
    \mathbf{D}^{\mathrm{mesh}}_m
    \|_{1}
    \Bigr],
\end{align}
where $\tilde{\mathbf{I}}_m$ and $\tilde{\mathbf{D}}_m$ are the RGB and depth rendered from the current 3DGS, and $\lambda_{\mathrm{r}}$, $\lambda_{\mathrm{s}}$, $\lambda_{\mathrm{p}}$, and $\lambda_{\mathrm{d}}$ are loss weights. 
The RGB reconstruction and DSSIM losses encourage the optimized 3DGS to match the generated observations. 
The LPIPS term helps reduce perceptual inconsistencies inherited from the generative model~\cite{cat3d}, while the depth term encourages the reconstructed geometry to remain aligned with the input meshes~\cite{worldmesh}.

\begin{figure*}[t]
    \centering
    \includegraphics[width=\textwidth]{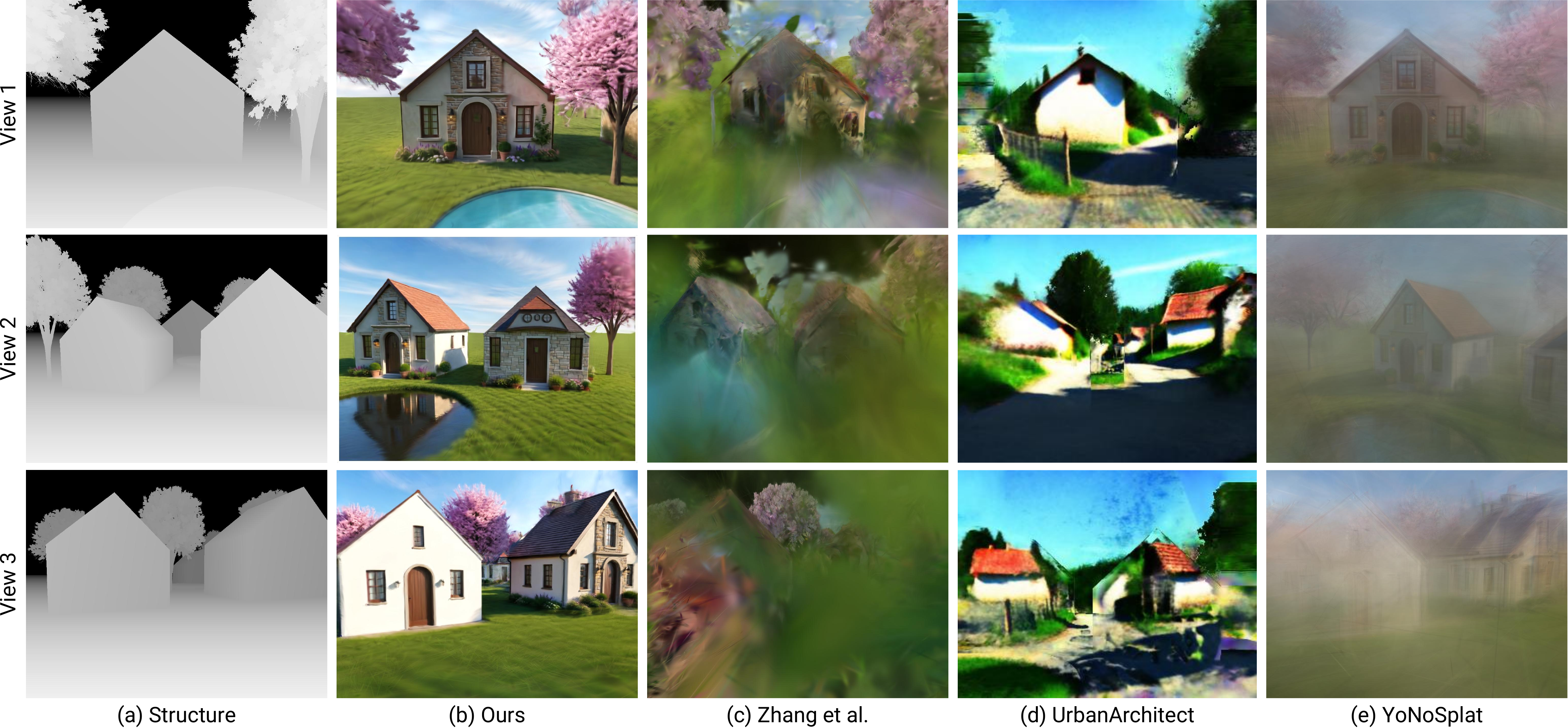}
    \caption{
    Qualitative comparison with baseline methods, including Zhang et al.'s method~\shortcite{continuous}, UrbanArchitect~\cite{urbanarchitect}, and YoNoSplat~\cite{yonosplat}.
    }
    \label{fig:cmp}
\end{figure*}
\begin{figure*}[t]
    \centering
    \includegraphics[width=\textwidth]{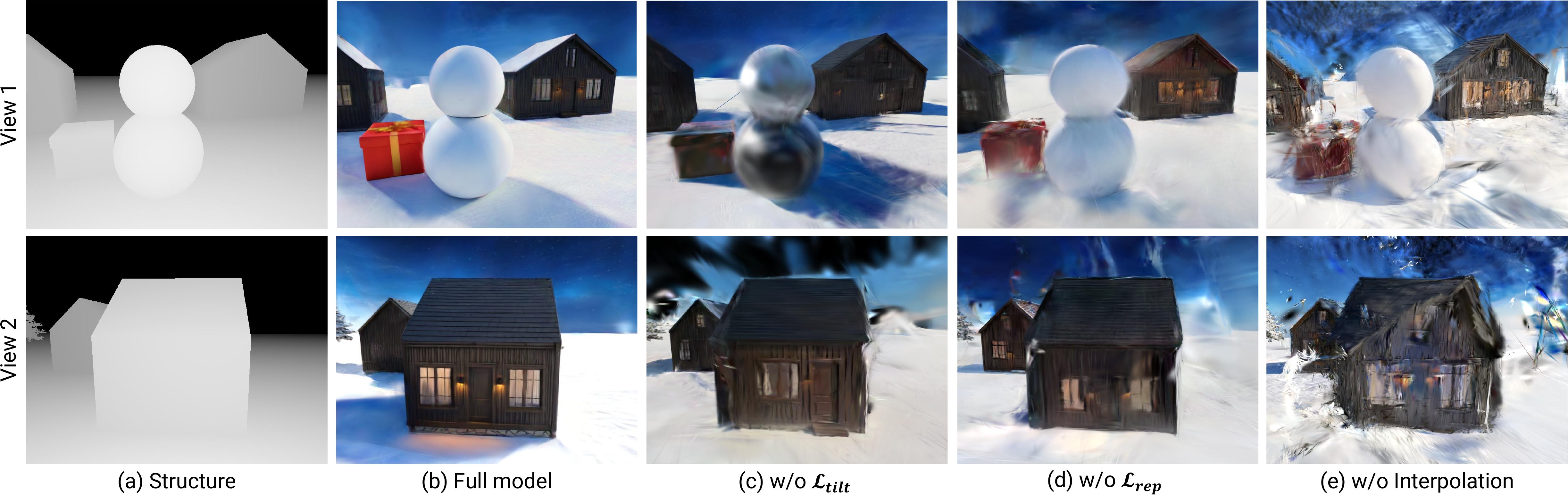}
    \caption{
    Qualitative ablation comparison of our full model and its variants, including configurations without tilt loss $\mathcal{L}_{tilt}$, repulsion loss $\mathcal{L}_{rep}$, and anchor-view interpolation, respectively. % \rt{scene will be replaced}
    }
    \label{fig:ablation_qual}
\end{figure*}
\begin{figure*}[t]
  \begin{center}
    \includegraphics[width=\textwidth]{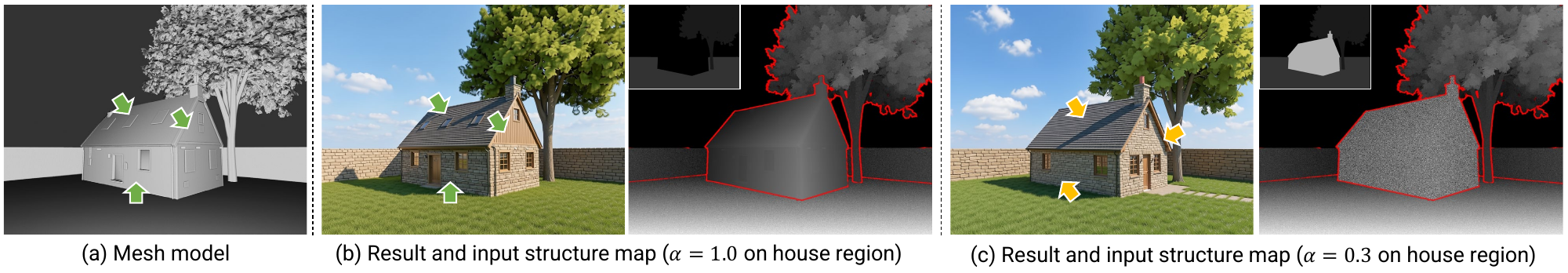}
  \end{center}
  \caption{
  Effect of geometry-adherence control. Given the same fine-detailed input geometry, larger $\alpha$ makes the generated object adhere more closely to the input structure map, while smaller $\alpha$ relaxes the geometry constraint.
  }
  \vspace{-3mm}
  \label{fig:alpha}
\end{figure*}

\begin{table*}[t]
    \centering
    \caption{
        Quantitative comparison with various baselines, including UrbanArchitect~\cite{urbanarchitect}, YoNoSplat~\cite{yonosplat}, and Zhang et al.'s method~\shortcite{continuous}, as well as ablation variants removing $\mathcal{L}_{\mathrm{tilt}}$, $\mathcal{L}_{\mathrm{rep}}$, or video interpolation. 
        Bold and underline indicate the best and second-best scores, respectively.
    }
    \vspace{-3mm}
    \renewcommand\tabcolsep{15pt}
    \resizebox{1\linewidth}{!}{
    \begin{tabular}{ccc|cc|ccc}
    \toprule
    \multicolumn{3}{c|}{Configuration}
    & \multicolumn{2}{c|}{Visual Quality}
    & \multicolumn{3}{c}{Structural Fidelity} \\
    View Scheduling
    & Multi-view Generation
    & 3D Optimization
    & CLIP Aesthetic$\uparrow$ 
    & MUSIQ$\uparrow$
    & PSNR-D$\uparrow$
    & Chamfer Distance$\downarrow$
    & F-score$\uparrow$ \\
    \hline\hline

    Our Module % UrbanArchitect
    & \multicolumn{2}{c|}{UrbanArchitect}
    & 4.360
    & 41.905
    & 14.521
    & 264.265
    & 0.00001 \\

    Our Module % YoNoSplat
    & Our Module
    & YoNoSplat
    & 3.413
    & 39.096
    & 15.796
    & 46.664
    & 0.00282 \\

    Zhang et al. % Zhang
    & Our Module
    & Our Module
    & 3.412
    & 33.782
    & 14.114
    & 39.173
    & 0.00990 \\

    \hline
    W/o $\mathcal{L}_{\mathrm{tilt}}$ % no_pitch
    & Our Module
    & Our Module
    & 5.066
    & 45.454
    & 20.647
    & \underline{20.100}
    & 0.01258 \\

    W/o $\mathcal{L}_{\mathrm{rep}}$ % no_repulse
    & Our Module
    & Our Module
    & \underline{5.987}
    & 49.588
    & \underline{21.531}
    & 21.061
    & \underline{0.01398} \\

    Our Module % no_interpolation
    & W/o interpolation
    & Our Module
    & 5.143
    & \textbf{59.653}
    & 19.082
    & 23.989
    & 0.00831 \\

    \hline
    \rowcolor{yellow!30}
    Our Module % Ours
    & Our Module
    & Our Module
    & \textbf{6.194}
    & \underline{54.474}
    & \textbf{21.974}
    & \textbf{19.255}
    & \textbf{0.01399} \\

    \bottomrule
    \end{tabular}
    }
    \label{tab:main}
    \vspace{-3mm}
\end{table*}

\section{Experiments}

In this section, we evaluate \ours through comprehensive experiments. 
Implementation details, including hyperparameters and thresholds, as well as additional comparisons, ablations, and analyses, are provided in the supplemental document.

\subsection{Scene Generation Results}
\cref{fig:qual} shows scene generation results using \ours across diverse geometry layouts, where each scene is guided by a varying set of identity images ranging from four to seven. 
The generated scenes faithfully follow the input 3D layouts, preserving the arrangement and shapes of scene elements.
At the same time, \ours successfully transfers appearance cues from the identity images to the corresponding elements, including sky, ground, trees, buildings, and walls. 
Moreover, the rendered images from cameras A--D further show that the generated 3DGS representations can be consistently rendered from arbitrary viewpoints that are not included in the generated observations, while maintaining high visual quality.

\subsection{Baseline Comparisons}
% \di{ this narrative "3 stages that enable xx baselines" is problematic. see if this is better "Our framework consists of three stages. To evaluate its effectiveness, we derive representative baselines by replacing individual stages with existing methods where applicable, enabling comparison across different generation paradigms."} 
Our framework consists of three stages. To evaluate its effectiveness, we derive representative baselines by replacing individual stages with existing methods where applicable, enabling comparison across different generation paradigms.
First, we compare to UrbanArchitect~\cite{urbanarchitect}, which follows an SDS-based optimization~\cite{dreamfusion} paradigm for scene generation. Since UrbanArchitect requires camera poses, we provide the camera poses obtained from our view scheduling algorithm. Second, we compare to YoNoSplat~\cite{yonosplat}, which follows a feedforward reconstruction paradigm. For this baseline, we provide our generated multi-view images with their corresponding camera poses, from which YoNoSplat reconstructs a 3DGS representation. 
In addition, to evaluate the effectiveness of our view scheduling algorithm, we compare it with a drone path-planning method~\cite{continuous}. This method is conceptually related to our task, as it also selects reconstruction-friendly viewpoints and computes trajectories from an input proxy mesh of the scene.

\begin{figure}[t]
  \begin{center}
    \includegraphics[width=1.0\linewidth]{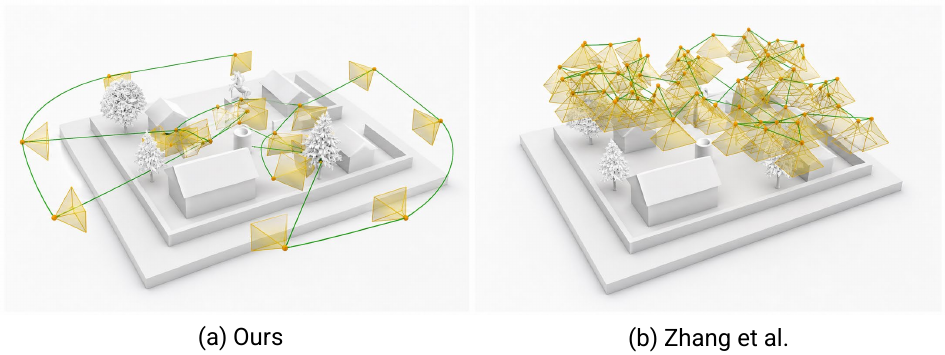}
  \end{center}
  \vspace{-3mm}
  \caption{
    View scheduling comparison with Zhang et al.'s method~\shortcite{continuous}.
  }
\vspace{-5mm}
  \label{fig:path}
\end{figure}

To measure visual quality, we use CLIP Aesthetic~\cite{clip-a}, and MUSIQ~\cite{musiq} on test frames.
For each scene, we construct a continuous camera trajectory by using the anchor-view nodes as control points of a B\'ezier curve, resulting in approximately 400--700 test frames.
These test frames do not overlap with the views used for optimization.
For structural fidelity, we measure scale-invariant PSNR on depth maps (PSNR-D)~\cite{videofrom3d}.
We also report Chamfer Distance and F-score with a distance threshold of 0.1.
To compute these metrics, we uniformly sample 100,000 points from the input mesh and compare them with the centers of the optimized Gaussian points.
For UrbanArchitect, which is based on a NeRF representation, we first extract a density field, reconstruct a mesh using marching cubes, and then uniformly sample points from the reconstructed mesh.
% \kkw{Testing other metrics}

For the test dataset, we construct 9 types of outdoor 3D scene layouts, consisting of either manually created coarse geometry or free assets.
Each scene contains a varying number of objects, ranging from 9 to 16.
From these scene models, we generate 10 different scenes for each baseline and use them for comparison.

\cref{fig:cmp} presents qualitative comparisons with the baseline methods.
Since Zhang et al.'s method targets aerial drone path planning, the resulting views tend to be closer to top-down views, as shown in \cref{fig:path}.
When depth maps are rendered from coarse scene geometry under these viewpoints, object shapes and camera poses can become ambiguous, which often leads to view generation failures.
Also, relying on such views degrades rendering quality at typical eye-level viewpoints.
UrbanArchitect suffers from blurry appearances and missing textures, as it relies only on SDS-based optimization with 2D image priors for 3D scene generation.
YoNoSplat is trained on real captured images and therefore suffers from a domain gap when applied to generated views.
Moreover, its applicability is limited to fewer than 100 input views, making it less effective for reconstructing large outdoor scenes from hundreds of generated observations.
\cref{tab:main} demonstrates the effectiveness of our method, showing that it outperforms the baselines across all metrics.

\subsection{Ablation \& Analysis}

\begin{figure}[t]
  \begin{center}
    \includegraphics[width=1.00\linewidth]{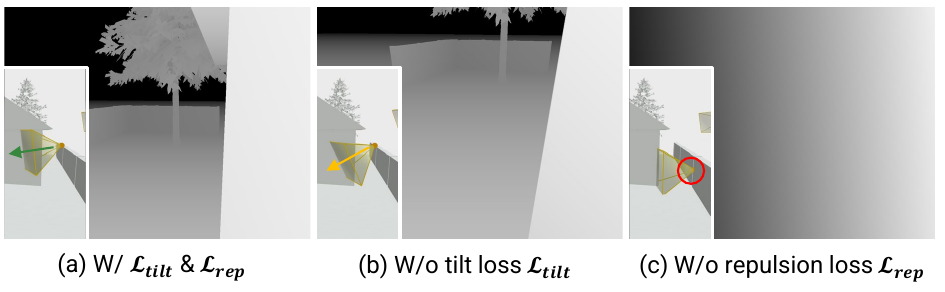}
  \end{center}
  \vspace{-5mm}
  \caption{
  Comparison of optimized camera poses and corresponding depth maps with and without the tilt and repulsion losses.
  }\label{fig:ablation_loss}
\end{figure}

\paragraph{Ablation on tilt loss $\mathcal{L}_{\mathrm{tilt}}$.}
\cref{fig:ablation_loss}(b) shows an example of camera pose optimization without the tilt loss. As indicated by the yellow arrow, the optimized camera direction is tilted toward the ground. This is because the ground mesh often accounts for a large fraction of the sampled surface points, which biases the optimized camera poses toward the ground plane. 
Consequently, the optimized views may fail to cover the full visible scene content, leading to incomplete appearances, as shown in \cref{fig:ablation_qual}(c).

\paragraph{Ablation on repulsion loss $\mathcal{L}_{\mathrm{rep}}$.}
\cref{fig:ablation_loss}(c) shows an optimization example without the repulsion loss.
As indicated by the red circle, the optimized cameras can move inside the input mesh. This results in invalid depth maps and collisions between camera paths and the input geometry, producing unreliable views and preventing these cameras from being connected to other nodes in the view graph. Consequently, some scene regions remain uncovered, leading to degraded reconstruction quality, as shown in \cref{fig:ablation_qual}(d).

\paragraph{Ablation on video interpolation.}
\cref{fig:ablation_qual}(e) shows the reconstruction result without video interpolation. In this setting, the available views are insufficient for stable 3DGS training, resulting in degraded rendering quality. Although this variant achieves the best scores on one visual quality metrics in \cref{tab:main}, these scores are likely affected by high-frequency artifacts favored by the metrics.

% \paragraph{Analysis on refinement loop.}
% \cref{fig:redundant_cam} shows how redundant cameras can arise during node initialization without the refinement loop.
% When a sampled view is blocked by geometry, the progressive initialization may re-sample nearby cameras with limited additional coverage.
% We address this by refining the initialized set through adding under-covered cameras, removing low-contribution cameras, and merging cameras with similar visibility.

\begin{figure}[t]
    \centering
    \includegraphics[width=\columnwidth]{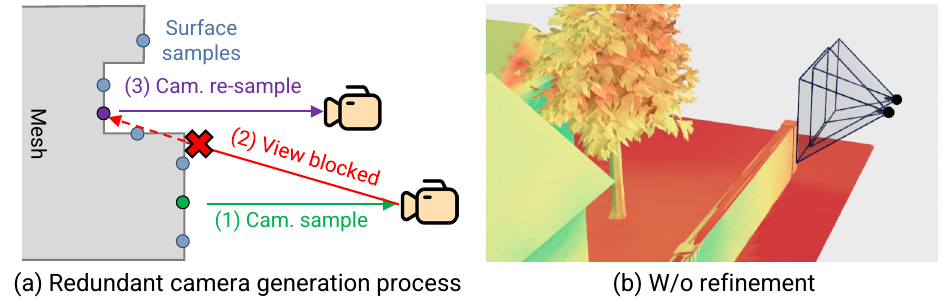}
    \vspace{-5mm}
    \caption{
    Effect of the refinement loop in node initialization. 
    (a) Illustration of how redundant cameras are generated when occluded views are not refined.
    (b) Node initialization result without the refinement loop. 
    }
    \vspace{-3mm}
    \label{fig:redundant_cam}
\end{figure}

\paragraph{Analysis on refinement loop.}
% \cref{fig:redundant_cam} illustrates how redundant cameras can arise during node initialization without the refinement loop. As shown in \cref{fig:redundant_cam}(a), nearby surface samples are expected to be covered by an initialized camera, but irregular input geometry can occlude even adjacent samples. The initialization process then generates another camera for these uncovered samples, producing a nearby camera with largely overlapping visibility. This leads to redundant cameras and poor scene coverage, as shown in \cref{fig:redundant_cam}(b). Our refinement loop mitigates this issue by removing low-contribution cameras, merging cameras with similar visibility, and adding new cameras for under-covered regions.
\cref{fig:redundant_cam}(a) illustrates how redundant cameras can arise during node initialization without the refinement loop.
Specifically, an initialized camera is expected to cover nearby surface samples, but irregular input geometry can occlude even adjacent samples. The initialization process then treats these occluded samples as uncovered and generates additional cameras near the existing one. Since these cameras are spatially close, they often observe largely overlapping regions, leading to redundant cameras and poor scene coverage (\cref{fig:redundant_cam}(b)). Our refinement loop mitigates this issue by removing low-contribution cameras, merging cameras with similar visibility, and adding new cameras from under-covered surface samples that have not yet initialized cameras.

\paragraph{Analysis on geometry-adherence control.}
\cref{fig:alpha} shows how the geometry-adherence parameter $\alpha$ controls the degree of geometric guidance during anchor-view generation.
Given detailed input geometry, setting $\alpha$ close to one makes the generated object closely follow the input structure map.
As indicated by the green arrows in \cref{fig:alpha}(b), fine geometric details such as the roof dormer and the positions of the door and windows are accurately reflected in the generated result.
In contrast, decreasing $\alpha$ relaxes the geometric constraint by adding noise to the structure map.
As indicated by the yellow arrows in \cref{fig:alpha}(c), the result follows the overall shape of the house while allowing more flexible appearance generation and weaker adherence to local geometric details.
This demonstrates that $\alpha$ provides controllable guidance between faithful geometry following and loose shape-level generation.

\begin{figure}[t]
  \begin{center}
    \includegraphics[width=1.0\linewidth]{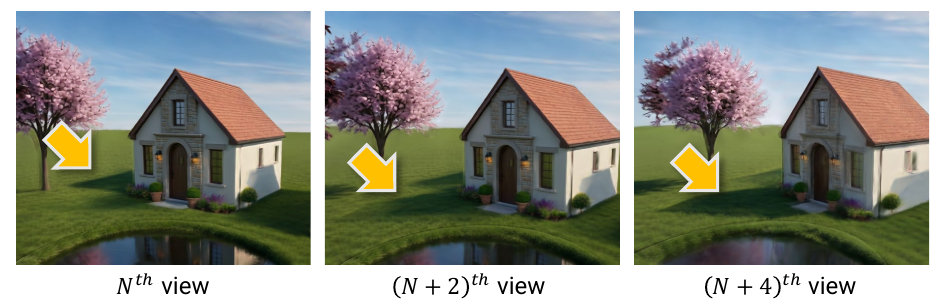}
  \end{center}
  \vspace{-3mm}
  \caption{
  Limitation of shadow consistency. The yellow arrows highlight shadows whose directions vary across views, leading to inconsistent appearance.
  }
  \vspace{-2mm}
  \label{fig:limit}
\end{figure}

\section{Conclusion}

In this paper, we introduced \ours, a geometry-conditioned framework for outdoor 3D scene generation without requiring explicit camera trajectories. By combining automatic view scheduling with an anchor-view-and-interpolation generation pipeline, our method enables high-quality 3DGS generation in large and unstructured outdoor scenes. As a first step toward object-level controllable 3D scene generation, our framework also provides control over object appearance and geometry adherence. Extensive experiments demonstrate its effectiveness across diverse scenarios.\\
\hspace*{1em}\textit{Limitations.}
Anchor-view generation may fail when a single view contains more than eight distinct object identities, due to the limited number of input images that the pretrained model can effectively incorporate. 
In addition, the lack of an explicit prior for global illumination can occasionally result in variations in shadow direction and size across anchor views, leading to shadow inconsistencies in the 3DGS output, as shown in \cref{fig:limit}.

\bibliographystyle{ACM-Reference-Format}
\bibliography{sections/reference}

% \documentclass[acmtog, anonymous, nonacm, review]{acmart}
%\documentclass[sigconf, anonymous, nonacm]{acmart}
% \acmSubmissionID{1151}

% \usepackage{booktabs} % For formal tables
% \usepackage[ruled]{algorithm2e} % For algorithms
% \usepackage{comment}

% % 추가한것
% \usepackage[capitalise]{cleveref}
% \usepackage{multirow}
% \usepackage{makecell}

% %나중에 지울것 한국어를 위한 패키지
% \usepackage{kotex}

% TOG prefers author-name bib system with square brackets
% \citestyle{acmauthoryear}

\clearpage
\pagebreak

\makeatletter
\renewcommand{\maketitle}{
  \begin{center}
    {\LARGE \bfseries \@title \par}
    \vspace{0.5em}
  \end{center}
}
\makeatother
\title{Supplemental Document}
\maketitle

\renewcommand{\algorithmcfname}{ALGORITHM}
\SetAlFnt{\small}
\SetAlCapFnt{\small}
\SetAlCapNameFnt{\small}
\SetAlCapHSkip{0pt}

% Metadata Information
\acmJournal{TOG}
\acmYear{2026}
\acmMonth{1}
\copyrightyear{2026}

% Document starts
% \input{var/var}
% \begin{document}

% Title portion
% \title{\ours{}: Geometry-Conditioned Outdoor 3D Scene Generation via View Scheduling with Object-Level Control}

\setcounter{section}{0}
\setcounter{figure}{0}
\setcounter{table}{0}

\renewcommand\thesection{S\arabic{section}}
\renewcommand\thefigure{S\arabic{figure}}
\renewcommand\thetable{S\arabic{table}}

% \maketitle

For a more detailed inspection of our results, please refer to the supplemental video.
We will release all code and datasets upon acceptance.
This supplementary material provides additional implementation details, statistics, and analyses for \ours{}.
It covers:
\begin{itemize}
    \item Input Geometry and Ground Mesh (\cref{sec:input_geometry})
    \item View Scheduling Details (\cref{sec:view_scheduling})
    \item Anchor-view Generation Details (\cref{sec:anchor_generation})
    \item Anchor-view Interpolation Details (\cref{sec:interpolation})
    \item 3DGS Optimization Details (\cref{sec:3dgs})
    \item Scene Statistics \& Latency (\cref{sec:statistics})
    \item Comparison of View Scheduling  (\cref{sec:scheduling_comparison})
    \item Ablation of Soft Visibility Score (\cref{sec:visibility_analysis})
    \item Ablation of 3DGS Training Loss (\cref{sec:3dgs_ablation})
\end{itemize}

\section{Input Geometry and Ground Mesh}
\label{sec:input_geometry}

The input geometry consists of controllable object meshes placed on a ground mesh.
We construct the ground mesh to be substantially larger than the object region, covering an area roughly ten times larger than the region occupied by the objects.
This prevents the boundary of a small ground mesh from appearing in the depth maps during view generation, which can otherwise cause artifacts such as floating patches of ground in the sky.

\section{View Scheduling Details}
\label{sec:view_scheduling}
The numerical parameters used by the main-paper view-scheduling formulation and the supplementary scheduling procedure are summarized in \cref{tab:schedule_hparams}.

\subsection{Surface Samples and Filtering}
We draw surface samples from the object meshes and from a bounded region of the ground mesh.
Because the ground mesh is intentionally larger than the object region, sampling the entire ground mesh would dominate the sample set with uninformative empty areas.
We therefore treat the XY bounding box of the object meshes as the region of interest and expand it by a fixed factor.
Ground samples are drawn only from the part of the ground mesh inside this expanded region.

Surface samples are generated by area-weighted barycentric sampling on mesh triangles followed by 3D Poisson-disk filtering~\cite{poissondisk}, which keeps the accepted samples approximately uniform over the mesh surface.
For a sampled surface area $A$ and surface-sample spacing $h$, we draw approximately $\lceil A / h^2 \rceil$ candidate samples before filtering.
The Poisson-disk minimum distance is set to the same spacing $h$.

After drawing surface samples, we remove invalid samples whose normals are immediately occluded.
Because the input object arrangement is user-specified, meshes can be placed in arbitrary contact configurations; for example, the floor surface of a house may directly touch the ground mesh.
Such contact regions can produce surface samples whose outward normal is immediately blocked by nearby geometry, making them uninformative for camera placement.
We therefore apply a normal-clearance filter that removes surface samples whose outward normal immediately intersects nearby geometry.

\begin{algorithm}[t]
\caption{Node initialization}
\label{alg:node_init}
\KwIn{Surface samples $\mathcal{P}=\{(\mathbf{p}_n,\mathbf{n}_n)\}_{n=1}^{N_{\mathrm{p}}}$}
\KwOut{Initial anchor-view set $\mathcal{V}$}
\textbf{Stage 1: Identify camera-sampleable samples}\;
$\mathcal{A} \leftarrow \emptyset$\tcp*[r]{camera-sampleable sample indices}
\For{$n \leftarrow 1$ \KwTo $N_{\mathrm{p}}$}{
    $c_n \leftarrow$ candidate camera from $(\mathbf{p}_n,\mathbf{n}_n)$\;
    \If{$|\mathrm{pitch}(c_n)| < \theta_{\max}$ and $H(c_n;\mathcal{P})>\delta_{\mathrm{cand}}$}{
        $\mathcal{A} \leftarrow \mathcal{A}\cup\{n\}$\;
    }
}
\BlankLine
\textbf{Stage 2: Progressive initialization}\;
$\mathcal{V} \leftarrow \emptyset$\;
\While{$\{n\in\mathcal{A}\mid \bar{V}_n(\mathcal{V})<\delta_{\mathrm{vis}}\}\neq\emptyset$}{
    Sample $n^*$ from $\{n\in\mathcal{A}\mid \bar{V}_n(\mathcal{V})<\delta_{\mathrm{vis}}\}$\;
    $\mathcal{V} \leftarrow \mathcal{V}\cup\{c_{n^*}\}$\;
    $\mathcal{A} \leftarrow \mathcal{A}\setminus\{n^*\}$\;
}
\BlankLine
\textbf{Stage 3: Refinement}\;
\Repeat{$\mathcal{V}=\mathcal{V}_{\mathrm{prev}}$}{
    $\mathcal{V}_{\mathrm{prev}} \leftarrow \mathcal{V}$\;
    $(\mathcal{V},\mathcal{A}) \leftarrow \mathrm{Add}(\mathcal{V},\mathcal{A},\mathcal{P})$\;
    $\mathcal{V} \leftarrow \mathrm{Remove}(\mathcal{V},\mathcal{P})$\;
    $\mathcal{V} \leftarrow \mathrm{Merge}(\mathcal{V},\mathcal{P})$\;
}
\BlankLine
\Return{$\mathcal{V}$}\;
\end{algorithm}

\subsection{Node Initialization}
\cref{alg:node_init} summarizes our node initialization procedure, which constructs an anchor-view set from the filtered surface samples before continuous pose optimization.
We use $\mathcal{P}$ for the filtered surface-sample set and $\mathcal{V}$ for the anchor-view set, following the main paper notation.

The first stage identifies the subset of samples that can be used to initialize cameras.
We introduce $\mathcal{A}\subseteq\mathcal{P}$ as the camera-sampleable subset used only for selecting new camera locations.
For each surface sample $(\mathbf{p},\mathbf{n})\in\mathcal{P}$, we place a candidate camera along the outward normal $\mathbf{n}$ at the preferred distance $d_0$, shortening the distance when geometry blocks the normal ray.
We denote by $B_{c,n}\in\{0,1\}$ the visibility mask that is one when $\mathbf{p}_n$ lies inside the frustum of camera $c$ and is not occluded by the input geometry.
The visibility contribution of camera $c$ is
\begin{align}
    H(c;\mathcal{P})
    =
    \sum_{n=1}^{N_{\mathrm{p}}}
    B_{c,n} V_{c,n},
\end{align}
where $V_{c,n}$ is the soft visibility score defined in the main paper.
The sample is added to $\mathcal{A}$ only if the candidate camera has pitch below $\theta_{\max}$ and $H(c;\mathcal{P})$ is above $\delta_{\mathrm{cand}}$.
Samples outside $\mathcal{A}$ remain in $\mathcal{P}$ for coverage evaluation, but are not used to initialize cameras.

The second stage progressively constructs the initial anchor-view set.
Starting from an empty camera set, we repeatedly compute the aggregated visibility of all surface samples and add a camera from the under-covered samples in $\mathcal{A}$.
The under-covered set is defined by the initialization visibility threshold $\delta_{\mathrm{vis}}$.
The new camera is placed along the selected sample normal and oriented to look back at the sample.
This progressive initialization stops when no eligible under-covered sample remains.

The progressive procedure provides a coverage-oriented initial set, but it does not explicitly optimize the compactness of the camera set.
The third stage therefore refines the initialized camera set before continuous pose optimization.
It adds cameras for still under-covered samples in $\mathcal{A}$, removes cameras with small visibility contribution, and merges camera pairs with similar visibility vectors.
Merging uses cosine similarity between the visibility rows $(V_{i,n})_{n=1}^{N_{\mathrm{p}}}$ and combines poses by a visibility-weighted mean.
The refinement thresholds are denoted by $\delta_{\mathrm{dense}}$, $\delta_{\mathrm{remove}}$, and $\delta_{\mathrm{merge}}$.
The refinement loop stops when the camera set no longer changes.
After node initialization, the camera poses are continuously optimized using the objective defined in the main paper while keeping the number of cameras fixed.

\subsection{Edge Construction}
For each edge selected by the scheduler, we store a smooth camera trajectory rather than only the two endpoint cameras.
This is intended to mimic natural in-domain camera motion during video interpolation, where the camera usually moves along a gentle path around the scene instead of following a purely linear endpoint transition.
For an accepted edge $(v_i,v_k)$, let $\mathbf{q}_i$ and $\mathbf{q}_k$ be the endpoint camera centers, and let $\mathbf{f}_i$ and $\mathbf{f}_k$ be their forward directions.
We define the trajectory $\Gamma_{i,k}$ by a cubic Bezier curve $\mathbf{q}(t)$, $t\in[0,1]$, whose endpoints are $\mathbf{q}_i$ and $\mathbf{q}_k$.
The two interior control points are initialized along the chord between the endpoints and then offset away from a common focus region estimated from the endpoint viewing rays.
The offset increases with the angular difference between $\mathbf{f}_i$ and $\mathbf{f}_k$, while being capped relative to the endpoint distance.
Thus, small-baseline edges remain close to a straight path, whereas wider-baseline edges follow a smoother arc around the scene.
If the sampled curve intersects the mesh, we fall back to the straight endpoint segment.

\subsection{Direction Construction and Generation Order}
Let $\mathcal{E}$ denote the undirected edge set over the anchor views.
To obtain a directed acyclic generation graph, we assign each anchor view a canonical order index $\rho(v_i)\in\{1,\dots,N_{\mathrm{v}}\}$ and orient every undirected edge from the lower-order endpoint to the higher-order endpoint:
\begin{align}
    \mathcal{E}_{\rightarrow}
    =
    \left\{
    (v_i,v_k)
    \mid
    \{v_i,v_k\}\in\mathcal{E},
    \rho(v_i)<\rho(v_k)
    \right\}.
\end{align}
Because $\rho$ strictly increases along every directed edge, the resulting graph $\mathcal{G}_{\mathrm{gen}}=(\mathcal{V},\mathcal{E}_{\rightarrow})$ is a DAG.
The anchor-view generation order is then any topological ordering of $\mathcal{G}_{\mathrm{gen}}$.
For a target anchor view $v_k$, its parent views are $\operatorname{Pa}(v_k)=\{v_i\mid(v_i,v_k)\in\mathcal{E}_{\rightarrow}\}$ and are generated before $v_k$.

\begin{table}[t]
    \centering
    \caption{View-scheduling parameter index.}
    \label{tab:schedule_hparams}
    \small
    \renewcommand\tabcolsep{15pt}
    \resizebox{1\linewidth}{!}{
    \begin{tabular}{lll}
    \toprule
    Notation & Name & Value \\
    \midrule
    $h$ & Surface-sample spacing & 1.0 \\
    $d_0$ & Preferred camera distance & 35.0 \\
    $\theta_{\max}$ & Candidate camera pitch limit & 10 degrees \\
    $\delta_{\mathrm{cand}}$ & Candidate visibility threshold & 32.0 \\
    $\delta_{\mathrm{vis}}$ & Initialization visibility threshold & 0.001 \\
    $\delta_{\mathrm{dense}}$ & Densification visibility threshold & 0.1 \\
    $\delta_{\mathrm{remove}}$ & Removal visibility threshold & 80.0 \\
    $\delta_{\mathrm{merge}}$ & Merge similarity threshold & 0.1 \\
    $\delta_{\mathrm{shared}}$ & Shared visibility threshold & 0.007 \\
    $\beta$ & Frustum sigmoid sharpness & 5.0 \\
    $\lambda_{\mathrm{front}}$ & Front-facing weight & 1.0 \\
    $d_{\mathrm{safe}}$ & Camera-mesh safety margin & 6.0 \\
    $\lambda_{\mathrm{rep}}$ & Camera-mesh repulsion weight & 1.0 \\
    $\lambda_{\mathrm{tilt}}$ & Pitch regularization weight & 1.0 \\
    \bottomrule
    \end{tabular}}
\end{table}

\section{Anchor-view Generation Details}
\label{sec:anchor_generation}

To enable the anchor-view generation conditioning used by \ours{}, we synthesize a task-specific training dataset and fine-tune a pretrained diffusion model.
We describe the dataset synthesis process and the training setup below.

\begin{figure*}[t]
    \centering
    \includegraphics[width=\textwidth]{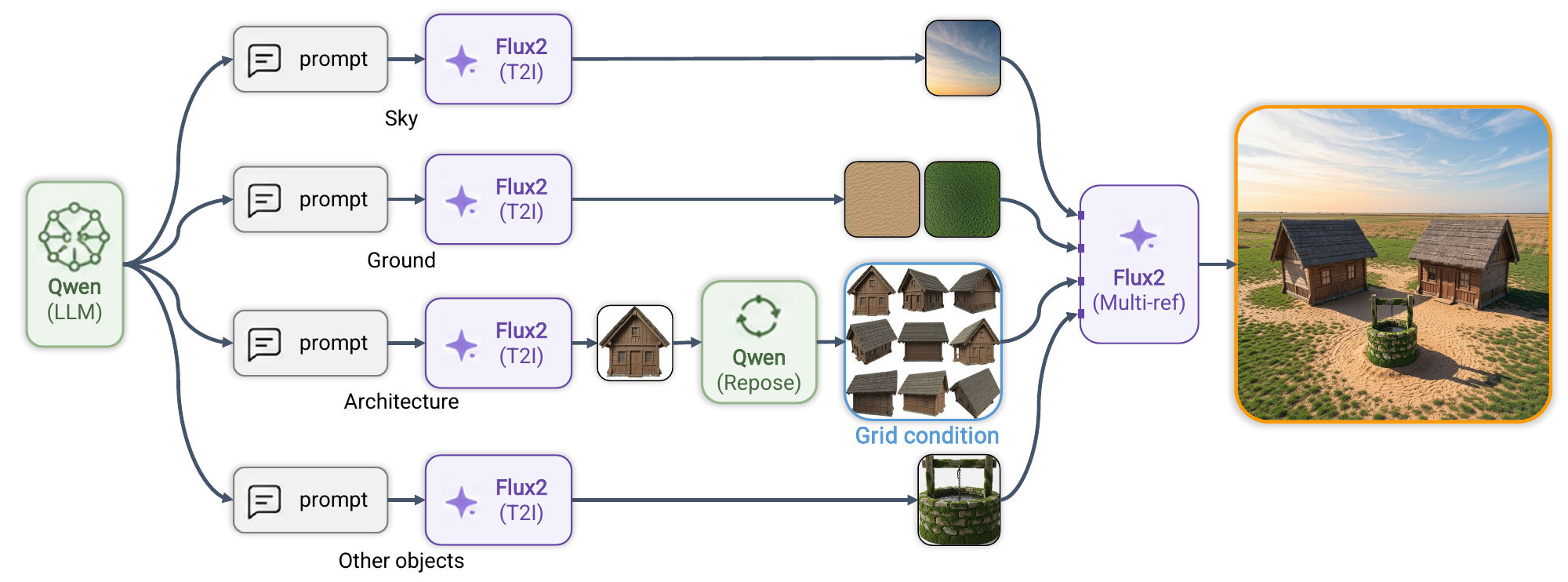}
    \caption{
    Dataset synthesis pipeline for identity images and anchor images.
    Object-specific identity prompts are generated first, then used to synthesize reusable identity images.
    For architecture components, we use a grid condition of multiple identity views to reduce pose bias.
    The resulting identity conditions are composed with sky and ground references to create multi-reference anchor-image training pairs.
    }
    \Description{A dataset synthesis pipeline diagram showing identity prompt generation, identity image synthesis, alternative identity views, and multi-reference anchor image synthesis.}
    \label{fig:dataset_pipeline}
\end{figure*}

\begin{figure}[t]
    \centering
    \includegraphics[width=\columnwidth]{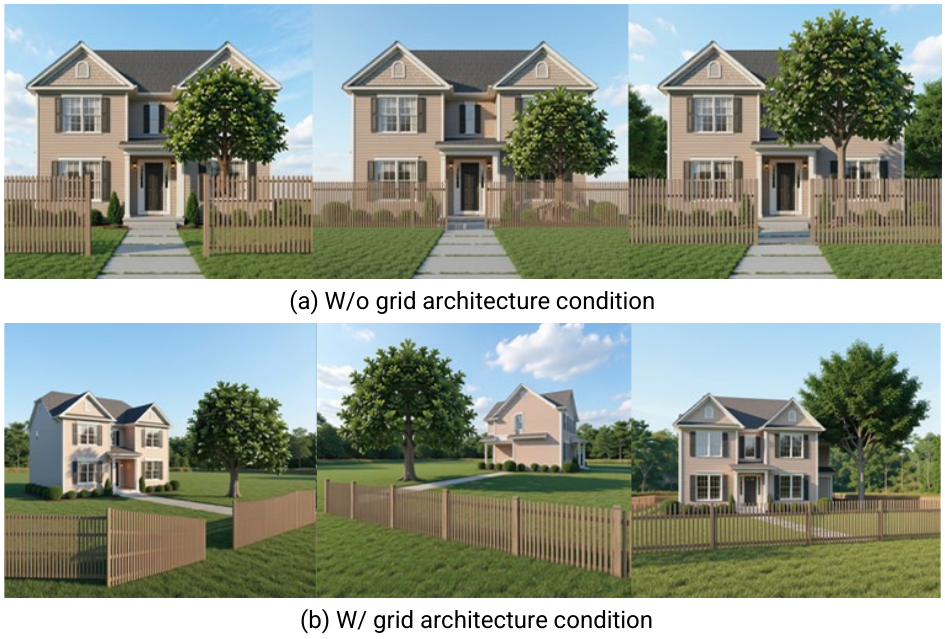}
    \caption{
    Motivation for grid conditioning of architecture components.
    The anchor-image synthesis model shows strong pose bias for buildings when conditioned on a single architecture reference.
    Under the same conditioning with different random seeds, a grid of multiple identity views exposes the generator to the same component from different viewpoints and reduces this bias.
    We use this grid condition only for architecture components.
    }
    \Description{A comparison illustrating that a single architecture identity reference can induce building pose bias, while a grid of multiple architecture identity views provides more viewpoint diversity.}
    \label{fig:grid_condition}
\end{figure}

\subsection{Dataset Synthesis}

Training requires paired examples that match the anchor-view generation input defined in the main paper.
Using the same notation, each synthetic training example consists of a target anchor image $\mathbf{x}_j$ and its conditioning tuple
\begin{align}
    \mathcal{C}_j
    =
    \Bigl(
        \mathbf{P}_j,\,
        \mathbf{O}_j,\,
        \hat{\mathbf{D}}_j,\,
        \mathbf{S}_j,\,
        \{(I_o,\mathbf{M}_{j,o})\}_{o\in\mathcal{O}_j},\,
        \tau_j
    \Bigr).
\end{align}
The following paragraphs describe how each component of $(\mathbf{x}_j,\mathcal{C}_j)$ is synthesized.

\subsubsection{Identity Images \& Anchor Image}

\cref{fig:dataset_pipeline} summarizes the dataset synthesis pipeline.
We first build an identity-image bank $\{I_o\}$ for the scene domain.
For each identity, we first generate an object-specific identity-image text prompt using Qwen3-8B\footnote{https://huggingface.co/Qwen/Qwen3-8B}, so the prompt describes the object category together with appearance attributes such as material, structure, and visual style.
For each object $o$, we use the FLUX.2 Klein base 9B\footnote{https://huggingface.co/black-forest-labs/FLUX.2-klein-9B} image diffusion model to generate an isolated identity image $I_o$ on a clean background, with prompts that emphasize a single centered object, realistic materials, and no surrounding scene clutter.
These identity images provide reusable appearance references for architecture and environment objects, while separate sky and ground images provide the background appearance references.
In total, the identity bank contains 349 identity images across 22 identity categories.

We then use the same FLUX.2 Klein model to synthesize full outdoor anchor images from the identity bank.
This synthesis is possible because FLUX.2 Klein supports multi-reference image generation: a single target anchor image can be conditioned on sky, ground, and multiple object references at once.
Each synthetic case samples a sky reference, one or two ground references, a camera-height description, and a visible object set $\mathcal{O}_j$ with identity images $\{I_o\}_{o\in\mathcal{O}_j}$.
The scene-generation prompt specifies the role of each input image, asks the model to preserve the referenced object identities and object counts, and requests one coherent full-frame outdoor photograph rather than a collage or contact sheet.
The result is the target anchor image $\mathbf{x}_j$, paired with metadata that records the semantic assignment of each reference image and is later used to construct the text prompt $\tau_j$.
This process yields approximately 17K identity-to-anchor-view training pairs.

For architecture components only, we use Qwen-Image-Edit-2511 with a multi-angle LoRA to generate multiple posed variants of the same identity and arrange them as a 3-by-3 grid condition.
We apply this grid conditioning only to architecture components because the anchor-image synthesis model exhibits a particularly strong pose bias for buildings: when given a single building reference, it tends to reproduce the reference viewpoint rather than adapting the component to the target scene.
As illustrated in \cref{fig:grid_condition}, under the same conditioning but with different random seeds, the grid condition exposes the generator to the same architecture component from several viewpoints, while ordinary environment objects continue to use a single identity reference.

\subsubsection{Semantic Mask}

From each synthetic anchor image $\mathbf{x}_j$, we estimate semantic masks for the visible sky, ground, and object regions.
We use Grounded-SAM2\footnote{https://github.com/IDEA-Research/Grounded-SAM-2} for this step. 
For every visible object $o\in\mathcal{O}_j$, the mask $\mathbf{M}_{j,o}$ indicates the image region occupied by that object in $\mathbf{x}_j$.
The identity-region conditions are therefore the paired inputs $\{(I_o,\mathbf{M}_{j,o})\}_{o\in\mathcal{O}_j}$, so the generator receives both the object appearance and the region where that appearance should be expressed.

\subsubsection{Partial Observation}

The pair $(\mathbf{P}_j,\mathbf{O}_j)$ represents the sparse RGB evidence available during sequential anchor-view generation.
During training, we obtain it by applying random pattern masks to the target anchor image $\mathbf{x}_j$, such as stripe-like masks or block-shaped masks, and then dropping observed pixels with Bernoulli noise thresholding.
The observed pixels form the partial observation $\mathbf{P}_j$, and the binary observation mask $\mathbf{O}_j$ records which pixels are retained.
For no-observation training cases, both $\mathbf{P}_j$ and $\mathbf{O}_j$ are set to zero.

\subsubsection{Structure Condition}

The pair $(\hat{\mathbf{D}}_j,\mathbf{S}_j)$ provides approximate geometric guidance without requiring exact target appearance.
For each synthetic anchor image, we estimate a clean depth map $\mathbf{D}_j$ with MoGe-2\footnote{https://github.com/microsoft/moge} and construct a per-pixel corruption strength map $\mathbf{S}_j$.
We then corrupt $\mathbf{D}_j$ according to $\mathbf{S}_j$ to obtain the structure condition $\hat{\mathbf{D}}_j$.
We also use the HED detector\footnote{https://github.com/s9xie/hed} to extract boundary cues that are overlaid on the corrupted depth condition.
This trains the generator to follow coarse structure while remaining robust to imperfect geometry cues.

\subsubsection{Prompt Condition}

The text condition $\tau_j$ is constructed by a rule-based template from the semantic assignments of the conditioning images.
The template first describes the fixed slot layout: image 1 contains $(\mathbf{P}_j,\mathbf{O}_j)$, image 2 contains $(\hat{\mathbf{D}}_j,\mathbf{S}_j)$, and each subsequent image contains an identity-region pair $(I_o,\mathbf{M}_{j,o})$.
It then appends object-specific clauses using the visible object names.
For example, when image 3 contains a house reference and image 4 contains a tree reference, the resulting prompt is:
\begin{quote}
\small
image1 stacks the downsampled partial observation on top of its observed-region mask.
image2 stacks the downsampled corrupted depth map on top of its per-pixel corruption strength.
image3 provides the house reference together with its region mask.
image4 provides the tree reference together with its region mask.
Generate one coherent scene that follows all image conditions, stays faithful to the depth in image2, and does not add extra objects.
\end{quote}

\subsection{Training}

We fine-tune FLUX.2 Klein with LoRA adapters on the transformer attention projections, while keeping the text encoder and VAE frozen.
All conditioning images are encoded by the VAE and concatenated as image tokens after the noisy target latent.
Separate image-token coordinates distinguish the target latent from each paired condition.
These paired conditions are $(\mathbf{P}_j,\mathbf{O}_j)$, $(\hat{\mathbf{D}}_j,\mathbf{S}_j)$, and the identity-region set $\{(I_o,\mathbf{M}_{j,o})\}_{o\in\mathcal{O}_j}$.
We use LoRA rank 128, alpha 128, and dropout 0.05.
The model is trained with AdamW using learning rate 1e-4 and weight decay 1e-4, with batch size 16 for 1500 iterations.

\subsection{Anchor-view Verification}

During sequential anchor-view inference, stochastic generation can occasionally place object content outside the intended semantic regions, especially when several object references are conditioned together.
To detect these failures, we evaluate a semantic overflow ratio for each generated anchor-view candidate.
We segment the generated image for the visible object categories, excluding sky and ground, and compare the union of the predicted object masks with the union of the target object regions in the conditioning semantic map.
The overflow ratio is the image-area fraction of predicted object pixels that lie outside the target object union.

If the overflow ratio is larger than 0.01, we discard the candidate and resample the same conditioning with the next random seed.
We evaluate at most 6 candidates in total and accept the first candidate whose overflow ratio is below the threshold.
If no candidate satisfies the threshold, we keep the candidate with the lowest overflow ratio.
This verification step only filters stochastic anchor-view failures; it does not change the scheduled view order or the conditioning inputs.

\section{Anchor-view Interpolation Details}
\label{sec:interpolation}

After the anchor images are generated, each scheduled edge is converted into a Wan2.1-VACE-14B interpolation input.
We use 49 frames for each edge.
The first and last frames are fixed to the two generated anchor-view images, while the 47 intermediate control frames are normalized depth renderings along the scheduled edge trajectory.
Video diffusion is run for 30 denoising steps.
The same fixed prompt is used for all edges:
\begin{quote}
\small
Generate a smooth, temporally coherent interpolation video between the first and last anchor-view frames.
Treat the first and last frames as strong appearance references, follow the intermediate control frames closely for geometry and camera motion, preserve the scene layout and object identity, keep textures and lighting consistent, and avoid flicker, deformation, new objects, or abrupt transitions.
\end{quote}

\begin{table}[t]
    \centering
    \caption{3DGS optimization parameter index.}
    \label{tab:3dgs_hparams}
    \small
    \renewcommand\tabcolsep{15pt}
    \resizebox{1\linewidth}{!}{
    \begin{tabular}{lll}
    \toprule
    Notation & Name & Value \\
    \midrule
    $\lambda_{\mathrm{r}}$ & RGB reconstruction weight & 0.8 \\
    $\lambda_{\mathrm{s}}$ & DSSIM weight & 0.2 \\
    $\lambda_{\mathrm{d}}$ & Metric-depth loss weight & 5.0 \\
    $\lambda_{\mathrm{p}}$ & Maximum LPIPS weight & 0.10 \\
    $T_{\mathrm{p}}$ & LPIPS start iteration & 5000 \\
    \bottomrule
    \end{tabular}}
\end{table}

\begin{table*}[!t]
    \centering
    \caption{
        Scene statistics and runtime summary of our method across different layouts.
        We report the number of objects, graph nodes, graph edges, and generated views for each layout.
        Runtime is measured in minutes and includes view scheduling, multi-view generation, 3DGS training, and the total pipeline runtime.
    }
    \renewcommand\tabcolsep{15pt}
    \resizebox{1\linewidth}{!}{
    \begin{tabular}{c|cccc|cccc}
    \toprule
    & \multicolumn{4}{c|}{Scene statistics}
    & \multicolumn{4}{c}{Latency (minutes)} \\
    Layout name
    & Object
    & Node
    & Edge
    & Views
    & View Scheduling
    & Multi-view Generation
    & 3DGS Training
    & Total \\
    \hline\hline

    Village
    & 16
    & 21
    & 22
    & 1099
    & 4.1
    & 316.8
    & 39.7
    & 360.6 \\
    
    Tribe Town
    & 16
    & 12
    & 12
    & 600
    & 17.0
    & 372.5
    & 37.3
    & 426.7 \\
    
    Desert
    & 12
    & 12
    & 12
    & 600
    & 8.0
    & 238.9
    & 36.2
    & 283.1 \\
    
    Cherry Blossom
    & 9
    & 10
    & 10
    & 500
    & 8.0
    & 185.2
    & 35.8
    & 229.0 \\

    Christmas
    & 11
    & 15
    & 17
    & 848
    & 2.3
    & 291.9
    & 43.7
    & 337.9 \\

    Ferris Wheel
    & 10
    & 12
    & 14
    & 698
    & 3.5
    & 286.7
    & 38.3
    & 328.5 \\

    Court Yard
    & 10
    & 6
    & 6
    & 300
    & 1.1
    & 115.6
    & 28.1
    & 144.8 \\

    Back Yard
    & 11
    & 11
    & 11
    & 550
    & 6.2
    & 165.3
    & 28.6
    & 200.1 \\

    Small Village
    & 9
    & 4
    & 4
    & 200
    & 9.6
    & 53.5
    & 23.7
    & 86.8 \\

    \bottomrule
    \end{tabular}
    }
    \label{tab:spec}
\end{table*}

\begin{figure*}[t]
    \centering
    \includegraphics[width=\textwidth]{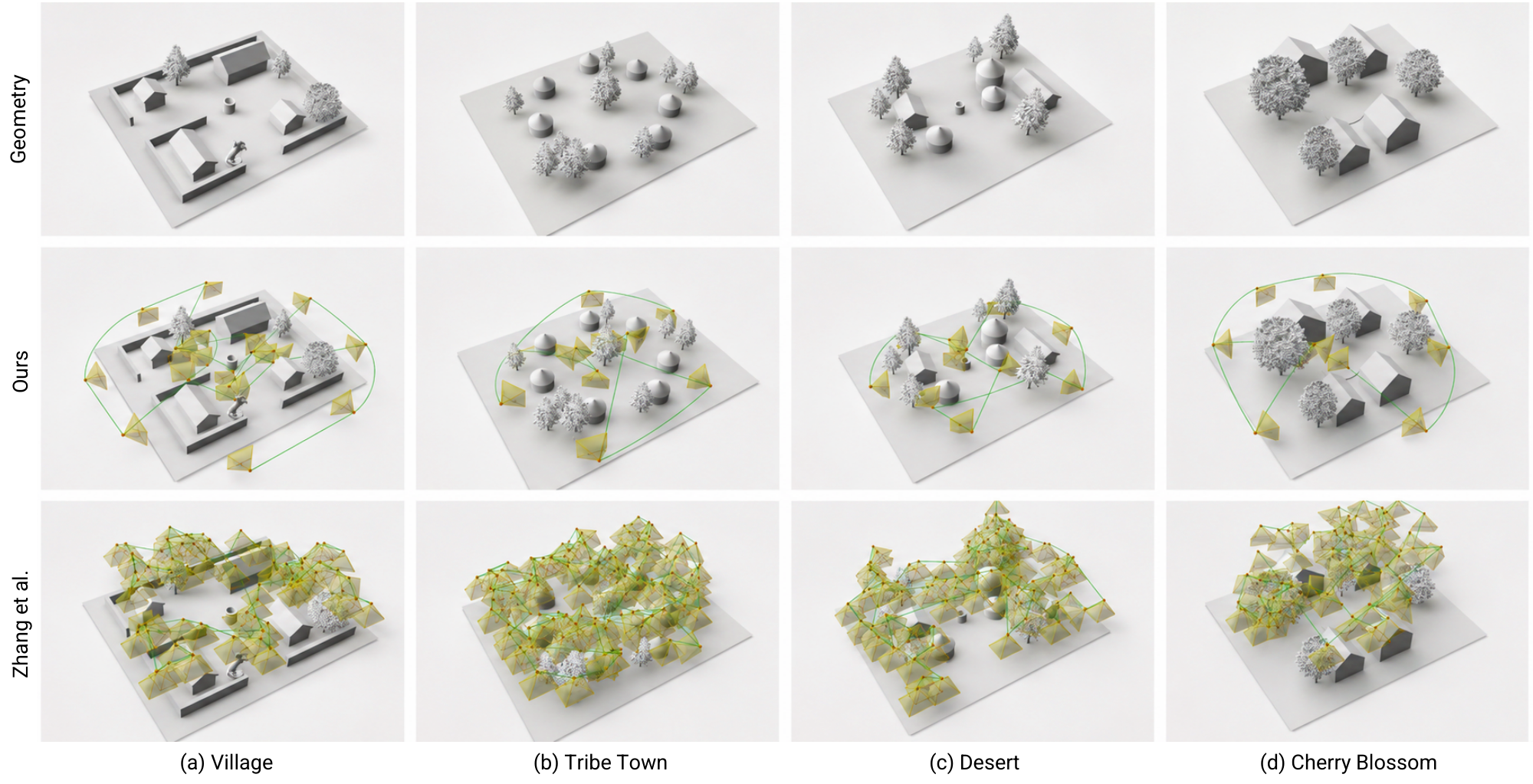}
    \caption{
    Qualitative comparison of view scheduling between our method and Zhang et al.~\shortcite{continuous}. Yellow frustums denote anchor views, and green edges denote interpolation trajectories.
    }
    \label{fig:supp_scheduling}
\end{figure*}

\section{3DGS Optimization Details}
\label{sec:3dgs}

We optimize the 3DGS model with the standard RGB reconstruction and DSSIM losses, plus metric-depth and LPIPS terms.
The numerical parameters used in this stage are summarized in \cref{tab:3dgs_hparams}.
Following CAT3D~\cite{cat3d}, the LPIPS coefficient is not fixed across all training views.
Following the main paper notation, let $\mathcal{J}$ denote the training-view index set.
Let $\mathcal{J}_{\mathrm{edge}}\subset\mathcal{J}$ denote the set of interpolated edge frames.
For an edge frame $m\in\mathcal{J}_{\mathrm{edge}}$ with frame index $r_m\in\{0,\dots,48\}$, we define its normalized position along the edge as
\begin{align}
    t_m = \frac{r_m}{48}.
\end{align}
We then use the midpoint-normalized edge weight
\begin{align}
    w_m = 4t_m(1-t_m),
\end{align}
which equals zero at the two endpoints and one at the midpoint.
The effective LPIPS weight is
\begin{align}
    \lambda_{\mathrm{p}}^{\mathrm{eff}}(m,\ell)
    =
    \begin{cases}
    \lambda_{\mathrm{p}} w_m,
    & m\in\mathcal{J}_{\mathrm{edge}},\ \ell\ge T_{\mathrm{p}},\\
    0,
    & \text{otherwise}.
    \end{cases}
\end{align}
This weighting gives zero LPIPS weight to anchor views and reaches its maximum at the midpoint of an interpolation edge.

\section{Scene Statistics \& Latency}
\label{sec:statistics}

\cref{tab:spec} reports the scene statistics and runtime of \ours across different layouts.
For each input layout, the table shows the number of objects, graph nodes, graph edges, and generated multi-view observations determined by our automatic view scheduling algorithm, together with the latency of each stage and the total pipeline runtime.
All experiments were conducted on NVIDIA A100-80G GPUs.
We used a single GPU for view scheduling and 3DGS training, while multi-view generation used two GPUs for parallel video inference.
On average, a single anchor-view generation takes approximately 4 minutes and a single interpolation takes approximately 7 minutes.
The reported anchor-view generation time includes retry attempts caused by verification failures.

\begin{figure*}[t]
    \centering
    \includegraphics[width=\textwidth]{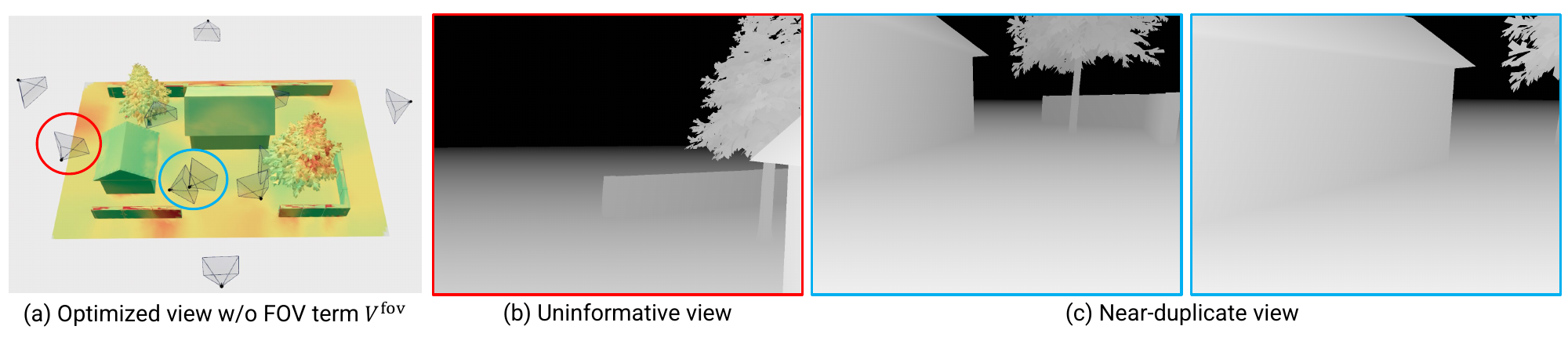}
    \caption{
Ablation example without the FOV term $\mathrm{V}^{\mathrm{fov}}$.
(a) Optimized camera poses without $\mathrm{V}^{\mathrm{fov}}$.
(b) View rendered from the red-circled camera in (a), which observes little informative scene content.
(c) Views rendered from the blue-circled cameras in (a), which are redundant with nearby views.
    }

    \label{fig:ablation_fov}
\end{figure*}

\begin{figure}[t]
    \centering
    \includegraphics[width=\columnwidth]{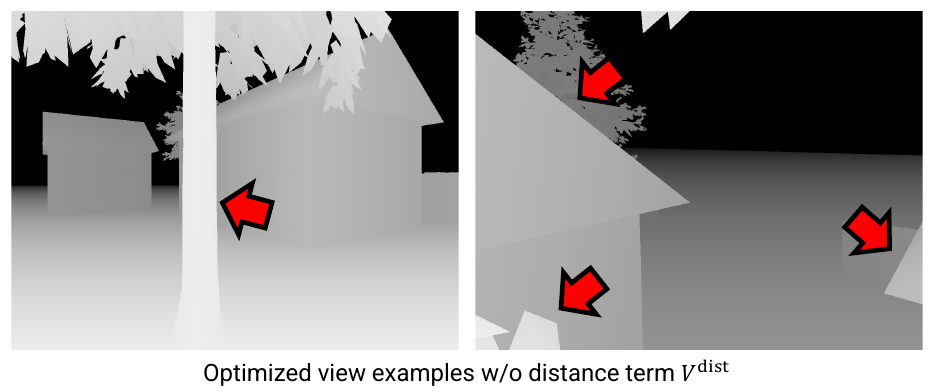}
    \vspace{-5mm}
    \caption{
Ablation examples without the distance term $\mathrm{V}^{\mathrm{dist}}$.
The rendered depth views show optimized cameras whose observations are largely blocked by nearby foreground geometry.
Red arrows indicate occluding objects placed close to the camera.
    }
    \label{fig:ablation_dist}
\end{figure}

\section{View Scheduling Comparison}
\label{sec:scheduling_comparison}

\cref{fig:supp_scheduling} provides qualitative comparisons of view scheduling across four input layouts.
Given the input geometry shown in the first row, \ours produces a sparse set of anchor views and interpolation trajectories that cover the scene structure, as visualized by the yellow frustums and green edges in the second row.
In contrast, Zhang et al.~\shortcite{continuous} generates substantially denser camera placements and trajectories, with many views biased toward overhead viewpoints.
This results in redundant observations and less effective coverage of object-level scene structures.
The comparison shows that \ours achieves more compact and structured view scheduling while maintaining coverage of the input layout.

\section{Analysis on Soft Visibility Score}
\label{sec:visibility_analysis}
We discuss the effect of each geometric term, including field-of-view, distance, and front-facing terms, used in our visibility score. 

\paragraph{Effect of field-of-view term.}
\cref{fig:ablation_fov} shows the result of removing the FOV term $\mathrm{V}^{\mathrm{fov}}$, which assigns visibility only to surface samples inside the camera frustum. 
This term is the most critical component because, without it, surface samples outside the camera view can receive the same visibility score as visible samples. 
As a result, the view optimization becomes almost random with respect to actual view coverage. 
This can produce uninformative views that observe little useful scene content, as shown in \cref{fig:ablation_fov}(b), or near-duplicate views that redundantly observe almost the same region, as shown in \cref{fig:ablation_fov}(c).

\paragraph{Effect of distance term.}
\cref{fig:ablation_dist} shows the results of removing the distance term $\mathrm{V}^{\mathrm{dist}}$. 
Without this term, surface samples are scored similarly regardless of their distance from the camera. 
Therefore, if a sufficient number of background surface samples fall within the view, the optimization can accept undesirable cameras even when a foreground object severely blocks the scene. 
As shown in \cref{fig:ablation_dist}, this often results in views where nearby geometry occludes the target regions, leading to poor and less useful observations for generation.

\begin{figure}
    \centering
    \includegraphics[width=\columnwidth]{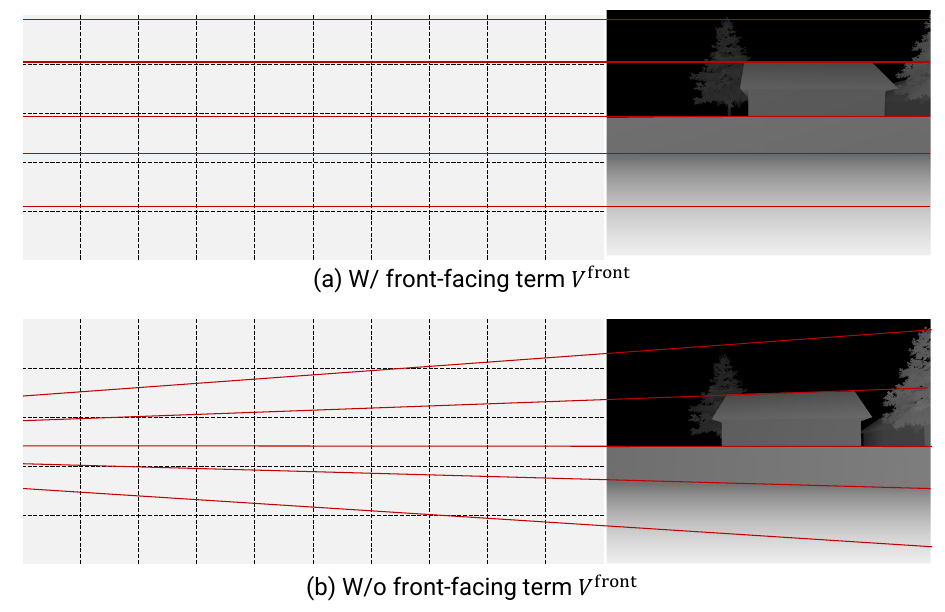}
    \vspace{-5mm}
    \caption{
Visualization of depth maps and perspective lines from optimized camera poses with and without the front-facing term $\mathcal{V}^{\mathrm{front}}$.
    }
    \label{fig:ablation_front}
\end{figure}

\paragraph{Effect of front-facing term.}
\cref{fig:ablation_front} shows the effect of the front-facing term $\mathrm{V}^{\mathrm{front}}$. 
This term encourages cameras to observe surfaces from a more frontal direction rather than from oblique angles. 
Such front-facing observations are generally more favorable for image generation because the target structure is clearer and less distorted in the rendered condition. 
In contrast, without the front-facing term, the optimized camera can observe surfaces at a slanted angle, which can degrade the quality of the generated views.

\begin{figure*}[t]
    \centering
    \includegraphics[width=\textwidth]{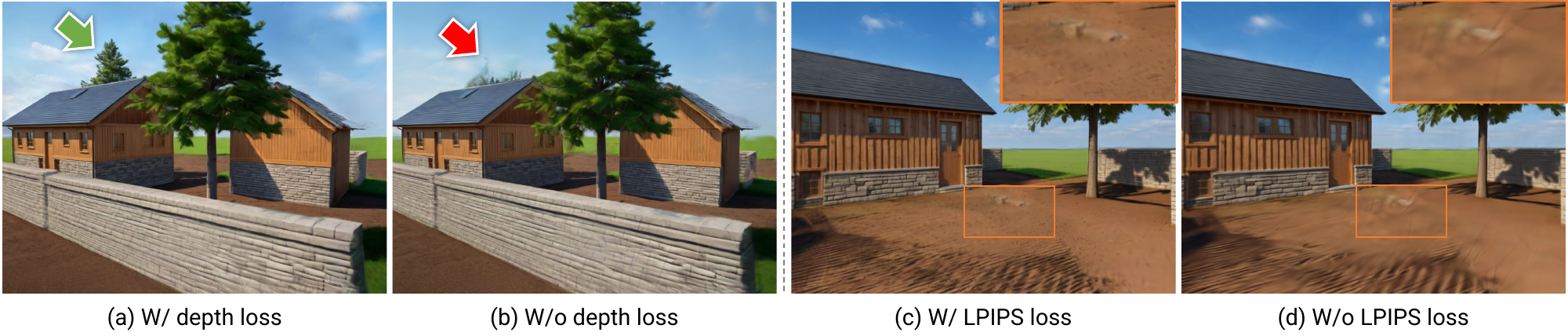}
    \caption{
Ablation examples of the 3DGS training losses.
(a,b) Results with and without the depth loss.
(c,d) Results with and without the LPIPS loss.
The arrows indicate structural differences, and the orange insets highlight texture differences.
    }

    \label{fig:ablation_3dgs}
\end{figure*}

\section{Ablation on 3DGS Training Loss}
\label{sec:3dgs_ablation}

\paragraph{Effect of depth loss.}
\cref{fig:ablation_3dgs}(a,b) shows the effect of the depth loss during 3DGS optimization.
Without the depth loss, the optimized Gaussians can drift away from the input geometry and produce floaters, as shown in \cref{fig:ablation_3dgs}(b).
These floaters can incorrectly appear in front of scene elements, occluding the tree in this example.
By enforcing consistency with the rendered depth from the input geometry, the depth loss suppresses such geometric artifacts and keeps the optimized 3DGS better aligned with the intended 3D structure, as shown in \cref{fig:ablation_3dgs}(a).

\paragraph{Effect of LPIPS loss.}
\cref{fig:ablation_3dgs}(c,d) shows the effect of the LPIPS loss during 3DGS optimization.
Without the LPIPS loss, high-frequency appearance details from the generated training views can be overly smoothed, resulting in blurry textures, as shown in \cref{fig:ablation_3dgs}(d).
This is particularly noticeable in the ground region highlighted by the orange inset.
By encouraging perceptual similarity to the generated views, the LPIPS loss helps preserve sharper and more detailed textures during 3DGS optimization, as shown in \cref{fig:ablation_3dgs}(c).

% \bibliographystyle{ACM-Reference-Format}
% \bibliography{sections/reference}

% \end{document}

\end{document}